\documentclass[twocolumn,trackchanges]{aastex63}

\usepackage{amsmath}
\usepackage{float}
\usepackage{url}
\usepackage{graphicx}
\usepackage{threeparttable} 
\usepackage{multirow}
\usepackage[figuresright]{rotating}

\shorttitle{S-Type stars in LAMOST DR10}
\shortauthors{Chen et al.}

\graphicspath{{./}{figures/}}

\begin{document}

\title{S-type stars from LAMOST DR10: classification of intrinsic and extrinsic stars}

\correspondingauthor{Yin-Bi Li \& A-Li Luo}
\email{* ybli@bao.ac.cn lal@nao.cas.cn   }

\author[0000-0001-8869-653X]{Jing Chen}
\affiliation{CAS Key Laboratory of Optical Astronomy, National Astronomical Observatories, Beijing 100101, China}
\affiliation{University of Chinese Academy of Sciences, Beijing 100049, China}

\author[0000-0001-7607-2666]{Yin-Bi Li$^{*}$}
\affiliation{CAS Key Laboratory of Optical Astronomy, National Astronomical Observatories, Beijing 100101, China}

\author[0000-0001-7865-2648]{A-Li Luo$^{*}$}
\affiliation{CAS Key Laboratory of Optical Astronomy, National Astronomical Observatories, Beijing 100101, China}
\affiliation{University of Chinese Academy of Sciences, Beijing 100049, China}
\affiliation{School of Information Management \& Institute for Astronomical Science, Dezhou University, Dezhou 253023, China}

\author[0000-0002-9279-2783]{Xiao-Xiao Ma}
\affiliation{CAS Key Laboratory of Optical Astronomy, National Astronomical Observatories, Beijing 100101, China}
\affiliation{University of Chinese Academy of Sciences, Beijing 100049, China}

\author[0000-0002-8913-3605]{Shuo Li}
\affiliation{CAS Key Laboratory of Optical Astronomy, National Astronomical Observatories, Beijing 100101, China}
\affiliation{University of Chinese Academy of Sciences, Beijing 100049, China}

\begin{abstract}

In this paper, we found 2939 S-type stars from LAMOST Data Release 10 using two machine-learning methods, and 2306 of them were reported for the first time. The main purpose of this work is to study how to divide S-type stars into intrinsic and extrinsic stars with photometric data and LAMOST spectra. Using infrared photometric data, we adopted two methods to distinguish S-type stars, i.e.,  XGBoost algorithm and color-color diagrams. We trained XGBoost model with 15 input features consisting of colors and absolute magnitudes of Two Micron All Sky Survey (2MASS), AllWISE, AKARI, and IRAS, and found that the model trained by input features with 2MASS, AKARI, and IRAS data has the highest accuracy of 95.52\%. Furthermore, using this XGBoost model, we found four color-color diagrams with six infrared color criteria to divide S-type stars, which has an accuracy of about 90\%. Applying the two methods to the 2939 S-type stars, 381 (XGBoost)/336 (color-color diagrams) intrinsic and 495 (XGBoost)/82 (color-color diagrams) extrinsic stars were classified, respectively. Using these photometrically classified intrinsic and extrinsic stars, we retrained XGBoost model with their blue and red medium-resolution spectra, and the 2939 stars were divided into 855 intrinsic and 2056 extrinsic stars from spectra with an accuracy of 94.82\%. In addition, we also found four spectral regions of Zr I (6451.6 \AA), Ne II (6539.6 \AA), $\mathrm{H_{\alpha}}$ (6564.5 \AA), and Fe I (6609.1 \AA) $\&$ C I (6611.4 \AA) are the most important features, which can reach an accuracy of 92.1\% when using them to classify S-type stars.

\end{abstract}


\section{Introduction} \label{sec:intro}

S-type stars are characterized by distinct ZrO molecular bands in the spectra, and their C/Os range from 0.5 to just below unity. S-type stars have been traditionally considered for a long time as intermediate red giants between M-type stars (C/O $<$ 0.5) and carbon stars (C/O $>$ 1.0) and in the evolution sequence of M-S-C at the asymptotic giant branch (AGB) phase \citep{1952ApJ...116...21M}. The spectra of S-type stars show a significant s-process elements overabundance, which are thought to be caused by the third dredge-up during the thermally pulsing asymptotic giant branch (TP-AGB) phase. However, this understanding was challenged when \cite{1988A&A...198..187J} found the S-type stars without Tc, which were later discovered that they are red giant branch (RGB; \cite{1992A&A...260..115J}) or early AGB \citep{1992ApJ...399..218B} in binary systems. The disappearance of Tc (a s-process element without stable isotopes with a half lifetime of about $\mathrm{2 * 10^5}$ yr) in these binary systems is due to its decay during the process of mass transfer. Hence, S-type stars can be classified into two different groups: Tc-rich S-type stars (intrinsic) and Tc-poor S-type stars (extrinsic; \cite{2022Univ....8..220V}). The intrinsic S-type stars with Tc and other s-process elements are caused by the third dredge-up in the TP-AGB stage, while the extrinsic S-type stars without Tc and their s-process elements' enrichment is caused by the mass transfer from the companion stars, which were formerly AGB stars and are white dwarf (WD) stars now. In addition, \cite{2020AA...635L...6S} recently discovered two peculiar bitrinsic S stars, which have properties of both the intrinsic and extrinsic stars.

The most direct way to distinguish intrinsic and extrinsic S-type stars is the existence of the Tc lines, which are very weak and blend with many other lines in medium- to low- resolution spectra and can be viewed in high-resolution spectra. Using high-resolution spectra (R = 30,000 - 60,000), with spectral coverage ranging from 4230 to 4270, \cite{1999A&A...345..127V} analyzed 70 S-type stars in the Henize sample \citep{1960AJ.....65..491H}, of which 41 were Tc-poor, and 29 were Tc-rich. \cite{2018AA...620A.148S} used HERMES (spectral resolution R=85,000) high-resolution spectra, and classified 16 S-type stars as extrinsic and 3 S-type stars as intrinsic based on Tc lines. However, detecting Tc lines to distinguish intrinsic and extrinsic S-type stars requires acquiring high-resolution, high signal-to-noise ratios (S/Ns) spectra, which is a big time and funding commitment. Fortunately, there are some indirect methods that can be used to distinguish intrinsic and extrinsic S-type stars, which were well summarized in \cite{1992IAUS..151..157J}, such as follows: 

\begin{itemize}
    \item $Periodic ~radial~ velocity~ (RV)~ variation$. There is a large dispersion in RVs for extrinsic S-type stars \citep{1990AJ.....99.1930B}, but intrinsic S-type stars can also display RV dispersion if they are in binary systems as in \cite{2020AA...635L...6S}. So it is not a foolproof method for distinguishing intrinsic and extrinsic S-type stars.
    \item $Direct~ observation~ of ~the ~hot~ companions~ with~ IUE$. If the WD companion star of the S-type star is hot enough, it can be detected in the ultraviolet region. However, both intrinsic and extrinsic S-type stars may have WD companions, so this method is also not foolproof.
    \item $High-excitation~ emission ~line~ of~ He~ I~ (10830 $\rm \AA$)$. \cite{1990AJ.....99.1930B} found that He I line has been detected in the spectra of all extrinsic S-type stars, and also in the spectra of one intrinsic S-type stars, so this method is also not foolproof.
    \item $Infrared~ (IR) ~excesses$. Stars in the evolution phase of TP-AGB have a greater mass loss, so the intrinsic S-type stars will have larger IR excesses and redder color than the extrinsic S-type stars.
\end{itemize}

Among them, the IR excesses method was widely used to distinguish intrinsic and extrinsic S-type stars, and a variety of IR colors were adopted. \cite{1993A&A...271..463J} investigated the IR properties for Stephenson's general catalog of S-type stars \citep{1984PW&SO...3....1S} using the Infrared Astronomical Satellite (IRAS) and the Two Micron All Sky Survey (2MASS) data. They found that the (K - [12], K - [25]) color-color diagram can be used to distinguish intrinsic from extrinsic S-type stars,  and at least 50\% stars in the Stephenson's general catalog are intrinsic S-type stars. \cite{1998A&A...333..613C} examined several color-color diagrams of 2MASS and IRAS by eye, and finally decided that K - [12] and [12] - [25] were the best criteria that could be used to distinguish between intrinsic and extrinsic S-type stars. \cite{2002A&A...387..129W} studied the near-IR features of 161 S-type stars, combined with IRAS photometric and low-resolution spectroscopy, and they found about 100 extrinsic S-type candidates. \cite{2006AJ....132.1468Y} selected about 150 extrinsic and 256 intrinsic S-type candidates from the General Catalogue of Galactic S stars \citep{1984PW&SO...3....1S} by combining IR data from 2MASS, IRAS, and the Midcourse Space Experiment. \cite{2019AJ....158...22C} used photometric data from 2MASS, IRAS, and the Wide Field Infrared Survey Explorer (WISE) and discovered 172 new intrinsic S-type stars. 

Traditionally, several feature indices, such as K - [12], [12] - [25], and [25] - [60], are used to distinguish between intrinsic and extrinsic S-type stars. However, these photometric features are selected by experience based on a small number of S-type spectra. Hence, it is not clear whether these features are the most effective ones to distinguish intrinsic and extrinsic S-type stars. In this work, we use a machine-learning method, eXtreme Gradient Boosting (XGBoost), to evaluate and quantify the importance of IR features used in this work, and find the most important features. In addition, using the intrinsic and extrinsic stars classified by the IR photometric data, we also try to classify them with the Large Sky Area Multi-Object Fiber Spectroscopic Telescope (LAMOST) medium-resolution spectra and study the most important spectral features to distinguish them.

In general, although the detection of Tc lines is the most accurate method to distinguish between intrinsic and extrinsic S-type stars, it is limited to the high-resolution and high S/Ns spectra. For MRS like LAMOST not cover the Tc lines or low-resolution spectra blended Tc regions, we tried to investigate classification methods using new IR colors and spectral features, which are indirect methods but suitable for large spectroscopic and photometric surveys.

This paper is organized as follows: In Section \ref{sec:search}, we found 2939 S-type stars from LAMOST Data Release (DR) 10, and identified them using three ways. In Section \ref{sec:distinguish}, two methods distinguishing intrinsic and extrinsic S-type stars with IR photometric data are presented. In Section \ref{sec:discuss}, with the help of the intrinsic and extrinsic S-type stars classified in section \ref{sec:distinguish}, we classified the 2939 S-type stars into intrinsic and extrinsic stars with both blue- and red-band MRS of LAMOST, and found the four most important spectral regions to distinguish intrinsic and extrinsic stars. Finally, the summary is presented in Section \ref{sec:summary}.

\section{S-type stars from LAMOST DR10} \label{sec:search}

\subsection{Data} \label{sec:lamost_spectra}

 LAMOST is a quasi-meridian reflecting Schmidt telescope with an effective aperture of 4 m, and implemented with 4000 fibers  \citep{1996ApOpt..35.5155W, 2004ChJAA...4....1S, 2012RAA....12..723Z, 2012RAA....12.1197C, 2012RAA....12.1243L, 2022Innov...300224Y}. Since 2018 October, LAMOST started the phase II survey, which contains both low- ($R$ $\sim$ 1800) and medium- ($R$ $\sim$ 7500) resolution spectroscopic surveys. From 2021 October to 2022 June, LAMOST DR10 collected 8,099,218 single-exposure spectra and 2,205,500 coadded spectra, which will be officially released in March of this year \footnote{\url{http://www.lamost.org/dr10/}}. The blue and red arms of the MRS spectra, which cover the wavelength range of 4950-5350 $\rm \AA$ and 6300-6800 $\rm \AA$, respectively, both have a resolution of 7500 at 5163 $\rm \AA$ and 6593 $\rm \AA$ \citep{2020arXiv200507210L}.

In our previous paper \citep[hereafter Paper I]{2022ApJ...931..133C}, 606 S-type stars were found from LAMOST MRS DR9 using the ZrO-band index greater than 0.25 as one of the sample selection criteria. However, applying the band index criterion before the machine learning would result in the loss of S-type stars as shown in Fig.6 of \cite{2017A&A...601A..10V}, which used the same method of calculating the band index as ours, and a fraction of S-type stars in this figure have ZrO indices lower than 0.25. In this work, only the machine-learning methods were used to find S-type stars from MRS of LAMOST DR10, and the 606 S-type stars from Paper I were treated as positive samples to search for more complete S-type samples.

\begin{figure*}
	\centering
        \includegraphics[width=1.0\textwidth]{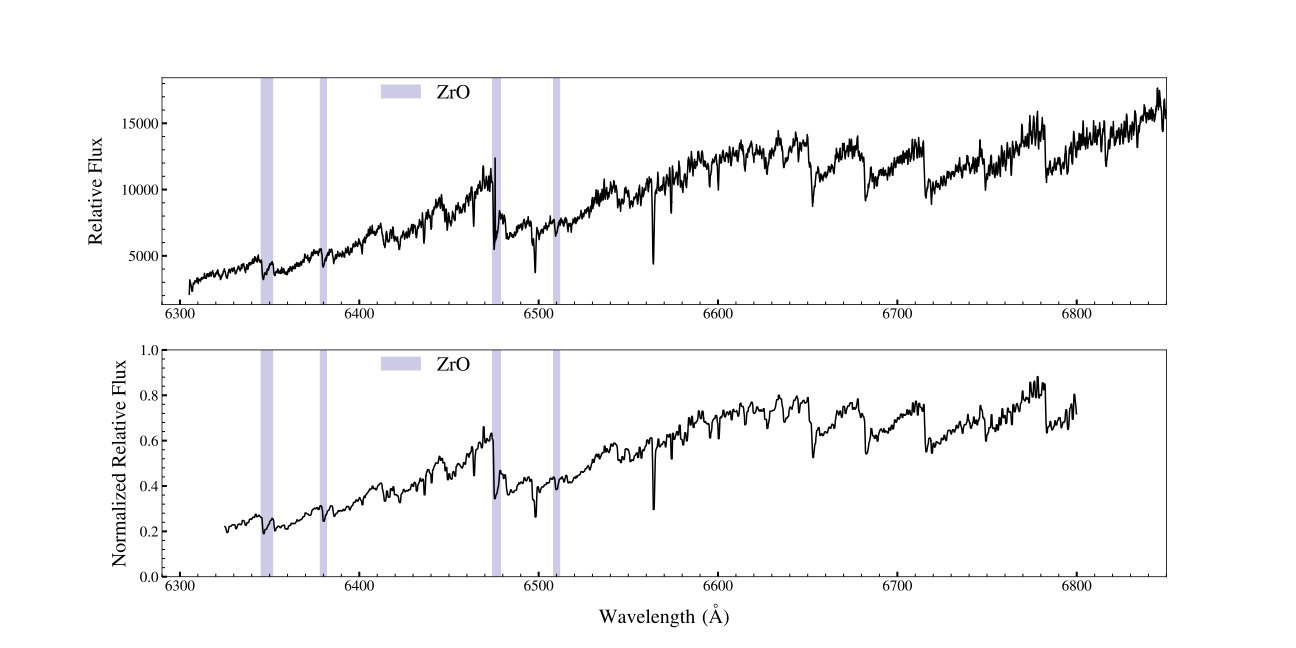}
	\caption{A medium-resolution spectrum example before (top) and after (bottom) spectral preprocessing. \label{Spectra_process}}
\end{figure*}

Before building the classifier, we preprocessed the MRS spectra first. The process of spectral preprocessing is the same as in Paper I: (1) only red-band spectra are shifted to the rest frame due to the wavelength range of ZrO band; (2) ``ormask," which denotes the quality of spectrum, equal to 0 is used to guarantee the spectra without problem; (3) interpolate the step size of each spectrum to 0.1 $\mathrm{\AA}$; (4) since there are missing values in the flux at wavelengths below 6325 $\mathrm{\AA}$ and above 6800 $\mathrm{\AA}$, we set the wavelength range between 6325 and 6800 $\mathrm{\AA}$; (5) normalize each spectrum by maximum flux. Besides, we found that cosmic rays occasionally pollute the ZrO bands, as in the top of Fig.\ref{Spectra_process}, so we smooth each spectrum with a 7 pixel median filter to remove noises. The parameters of the filters are set empirically so that the absorption lines in the spectra are not affected. Figure \ref{Spectra_process} shows a red-band LAMOST MRS spectrum before and after filtering. 

\subsection{Algorithm}

To find S-type stars more completely, we used two machine-learning methods, Support Vector Machine (SVM) and XGBoost, and combined their results together. 

SVM is a supervised learning algorithm, developed by \cite{1995ML...20....273} as an extension to nonlinear models of the generalized portrait algorithm. Here, we used the Scikit-Learn module implementation of a soft margin SVM in the Python programming language, and this implementation is both robust and fast. The SVM can be either linear or nonlinear. A linear classifier separates the examples by finding a hyperplane in the feature space, a nonlinear classifier separates the examples by a more complicated hypersurface, i.e. using kernels to remap the feature space into a higher-dimensional space. Whether in the linear classifier or in the nonlinear classifier, the SVM tries to find the separating hyperplane with the largest distance to the nearest training examples, i.e. it is a maximum margin classifier.

XGBoost is an implementation of gradient-boosted decision trees, a set of machine-learning techniques used for regression and classification \citep{2016arXiv160302754C}. Each decision tree learns the residual of the sum of the target value and the predicted value of all previous trees. Multiple decision trees make decisions together and finally add up the results of all trees as the final prediction result. XGBoost performs significantly faster and more accurately than similar gradient-boosted tree solutions. XGBoost includes regularization terms in the objective function to control the model complexity and therefore can reduce overfitting and improve the model generalization. It is also optimized for sparse input data, i.e., data with missing values.

\subsection{Search for S-type stars with the two algorithms from LAMOST DR10}
\label{sec:lamost_dr10_sstar}

There are 10,608 single-exposure and coadded red-band spectra for the 606 S-type stars in Paper I after removing spectra with flux absence and spectral wavelength ranges that cannot be interpolated between 6325 and 6800 $\rm \AA$. For spectra with low S/Ns and low quality (such as high S/Ns but discontinuous wavelength or strange spectral shapes), we removed them manually, and finally, 4585 spectra were remaining and treated as positive samples. To avoid the issue of the category imbalance in the training samples for machine-learning algorithms, we randomly selected the same amount (4585) of red-band MRS from LAMOST DR10 as the negative sample, covering all spectral types and removing contamination of S-type stars by eye. The positive and negative samples were randomly divided into the training and test data sets according to the ratio of 8:2, the training data were used to build the SVM and XGBoost classifiers, and the test data were used to evaluate the accuracies of classifiers. 
The four ZrO regions,  previously defined in Paper I, were treated as input features, and the accuracies of SVM and XGBoost models can be calculated as follows:
\begin{gather}\label{1}
    \mathrm{Acc = \frac{TP + TN}{TP+TN+FP+FN}},
\end{gather}
where TP, TN, FP, FN are the number of true positive, true negative, false positive, and false negative, respectively. The trained SVM and XGBoost models were applied to the test data, and their accuracies are 97.93$\%$ and 99.02$\%$ respectively, which are shown in Table \ref{tab:modelresult}. It should be pointed out that the accuracies in Sections 3.1.1 and 4.1 were also calculated by formula (1).

\begin{deluxetable}{ccccc}
\caption{Model Accuracy Using the Test Dataset and Predicted Number of SVM and XGBoost Models.}
\tablehead{
\colhead{Model} & \colhead{Accuracy} & \colhead{Predicted Number} & \dcolhead{\rm (S/N) > 5} & Same
}
\startdata
SVM & 97.93$\%$ & 299,120 & 119,696 & \multirow{2}{*}{80,738} \\
XGBoost & 99.02$\%$ & 266,120 & 99,116 & \\
\enddata
\end{deluxetable} \label{tab:modelresult}

After the above steps, our SVM and XGBoost prediction models were applied to all MRS red-band single-exposure and coadded spectra of LAMOST DR10, and only 20,309,356 spectra with ``fibertype = Obj" or ``fibertype = F-std" were used to search for S-type stars in this work. There are nearly 300,000 spectra classified as S-type stars by SVM and XGBoost, of which the low S/Ns spectra account for a large part, so we eliminated spectra with S/Ns less than 5. Finally, each model predicted about 100,000 spectra with S/Ns $>=$ 5, of which more than 80,000 spectra were predicted by both models as S-type stars; the specific numbers are listed in Table \ref{tab:modelresult}. We finally manually examined 138,074 spectra and found 24,060 spectra (corresponding 2929 stars) have ZrO bands, whose bandheads can be seen in Table 2 of Paper I. Additionally, 10 stars in Paper I were not included in this work, which can be attributed to the fact that only one loop was performed in this study, while multiple loops were conducted in Paper I for a more comprehensive search. 
So 2939 S-type stars were finally found in this work after adding the 10 stars.

\subsection{Further Identification of S-type stars}\label{sec:identify}

\begin{figure}[ht!] 
	\centering
        \includegraphics[width=0.5\textwidth]{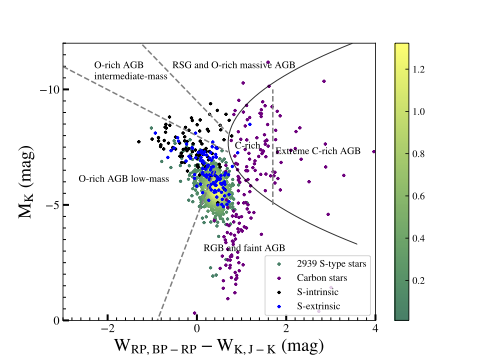}
	\caption{$M_{K} ~vs.~ W_{\mathrm{RP, BP-RP}} - W_{K, J-K}$, diagram of S-type stars. The green and purple dots denote 764 S-type stars and carbon stars (the ``C-N" type stars from \cite{2018ApJS..234...31L}), respectively. The black and blue dots show the intrinsic and extrinsic S-type samples from \cite{2022A&A...664A..45A}. The solid black curve represents the theoretical boundary between O-rich and C-rich AGB stars. The dashed lines denote the threshold that distinguishes different types of AGB stars. The color bar on the right indicates the probability density of S-type stars.\label{CMD}}
\end{figure}

To verify that these 2939 stars are indeed S-type stars, we used the method provided by \cite{2018A&A...616L..13L}, which is based on Gaia and 2MASS photometry, and can be used to distinguish RGB or faint AGB, low-mass, intermediate-mass, and massive O-rich AGB stars, and extreme C-rich AGB stars; the mass range of each group is related to its metallicity. \cite{2020A&A...633A.135A} has demonstrated that this method could provide a powerful tool for studying and identifying different spectral types of the AGB-phase stars, as shown in Fig.\ref{CMD}. The three dashed lines and the curve in this figure were defined by \cite{2018A&A...616L..13L} (see Table 1 from their work) for the long-period variable objects in the Large Magellanic Cloud, and we converted the apparent magnitude of $K_S$ to the absolute magnitude of $M_K$ by assuming the distance modulus of the Large Magellanic Cloud is 18.49 \citep{2004NewAR..48..659A}. In this diagram of $M_{K} ~vs.~ W_{RP, BP-RP} - W_{K, J-K}$, $M_K$ is the $K$ band absolute magnitude, and $W_{RP, BP-RP}$ and $W_{K, J-K}$ are reddening-free Wesenheit functions \citep{2005PASP..117..823S}, which were defined as follows:
\begin{gather}
    \mathrm{W_{RP, BP-RP} = G_{RP} - 1.3(G_{BP} - G_{RP})}, \\
    \mathrm{W_{K, J-K} = K - 0.686(J - K)},
\end{gather}
where $\mathrm{G}_{BP}$ and $\mathrm{G}_{RP}$ are the Gaia blue- and red-band magnitudes, respectively. We adopted the last three criteria of selection (A) in Appendix C of \cite{2018A&A...616A...2L}, as well as the quality cuts (i), (iv), and (v) of Section 4 in \cite{2019MNRAS.490..157M} to select 764 S-type stars with high-quality Gaia data, and their distributions on $M_{K} ~vs.~ W_{RP, BP-RP} - W_{K, J-K}$ are shown in Fig.\ref{CMD}. The green dots are the 764 S-type stars, and the 107 extrinsic and 90 intrinsic S-type stars from \cite{2022A&A...664A..45A} were plotted as the black and blue dots, respectively, for comparison. We can see that most of the 2939 S-type stars are indeed located in the O-rich AGB phase and are consistent with the distribution of known intrinsic and extrinsic S-type stars.

\begin{figure}[ht!] 
    \centering
    \includegraphics[width=0.5\textwidth]{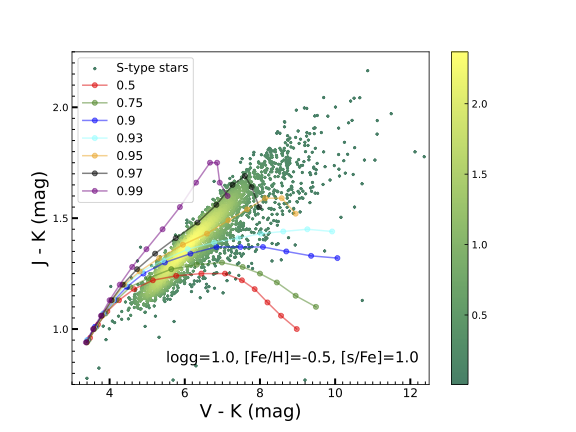}
    \caption{The $V - K$ and $J - K$ diagram for the 2920 S-type stars with 2MASS and Pan-STARRS 1 photometric data. The green dots represent S-type stars, and the seven color lines denote MARCS models with C/O = 0.5, 0.75, 0.9, 0.93, 0.95, 0.97, 0.99, respectively. The color bar on the right denotes the probability density of S-type stars.\label{MARCS_model}}
\end{figure}

\begin{figure}[ht!] 
    \includegraphics[width=0.45\textwidth]{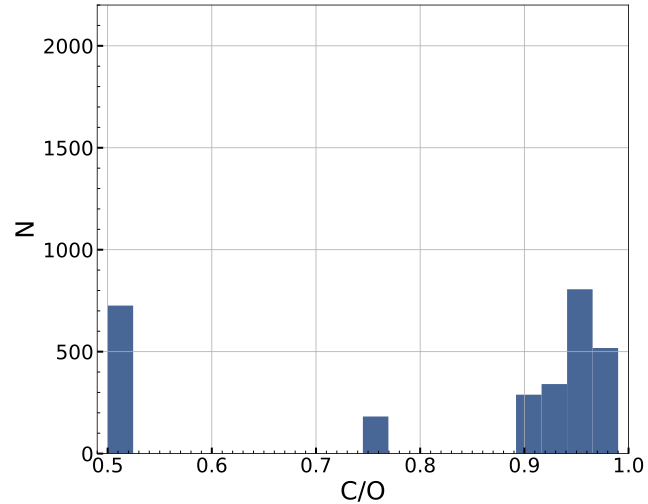}
    \caption{The C/O distribution for 2920 S-type stars estimated by the MARCS S-type stars model.\label{MARCS_CO}}
\end{figure}

As in Paper I, we used two methods to estimate C/O, i.e., the $(V - K, J - K)$ photometric indices and [C/Fe] and [O/Fe] from APOGEE. \cite{2017A&A...601A..10V} provided a grid of MARCS model atmospheres for S-type stars with $\mathrm{0.50\leq C/O \leq 0.99}$, and the $(V - K, J - K)$ color-color diagram constructed from the MARCS models was used to roughly estimate the C/Os of S-type stars in this work. We cross-matched the 2939 S-type stars with 2MASS and Pan-STARRS 1 \citep{chambers2019panstarrs1} in 3$''$, obtained $J$, $K$, $g$, and $r$ magnitudes of 2920 common stars, and their $V$ magnitudes were transformed from g and r magnitudes using the method in \cite{2006astro.ph.12034Z}. The extinctions of $J-$, $K-$, $g-$, and $r-$band magnitudes were estimated by using the Python package $\rm dustmaps$\footnote{\url{http://argonaut.skymaps.info}}, which presented a three-dimensional map of dust reddening based on Gaia parallaxes and stellar photometry from 2MASS and Pan-STARRS 1 \citep{2019ApJ...887...93G}. If given log $g$, [Fe/H], and [s/Fe], the C/Os can be estimated by $V - K$ and $J - K$ colors. However, the three parameters of the 2920 stars are unknown; we referred the log g, [Fe/H], and [s/Fe] of the 38 S-type stars in \cite{2018AA...620A.148S, 2019AA...625L...1S, 2020AA...635L...6S, 2021AA...650A.118S}, where their log $g$ is 0 or 1 dex, [Fe/H] ranges from -0.5 to 0.0, and [s/Fe] equals 0 or 1 dex, to roughly constrain the three parameters for the 2920 stars. Eight $(V - K, J - K)$ diagrams were thus constructed by the MARCS models with log $g$ = 0.0 or 1.0, [Fe/H] = -0.5 or 0.0, and [s/Fe] = 0.0 or 1.0, and were used to estimate the C/O ratios for the 2920 S-type stars, respectively. On these diagrams, the stars generally distribute consistent with the MARCS models, which indirectly indicates that they are likely S-type stars, the C/O ratios from each color-color diagram are all between 0.5 and 1.0 as expected, and most of them are greater than 0.9. We calculated the average values of log $g$, [Fe/H], and [s/Fe] of the 38 S-type stars (log $g$ = 0.8, [Fe/H] = -0.27, [s/Fe] = 0.7), and chose C/Os from the color-color diagram with log $g$ = 1.0, [Fe/H] = -0.5, and [s/Fe] = 1.0 (closest to the average values) as the final results listed in Table \ref{tab:catalog}. Figure \ref{MARCS_model} shows the selected $(V - K, J - K)$ diagram, and Fig.\ref{MARCS_CO} shows the C/O distribution of the 2920 S-type stars estimated from Fig.\ref{MARCS_model}. 

\begin{figure}[ht!] 
    \centering
    \includegraphics[width=0.5\textwidth]{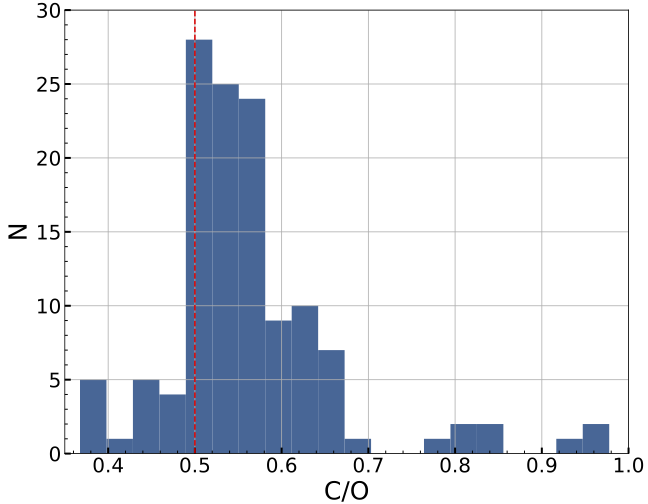}
    \caption{The C/O distribution for 132 S-type stars with C/Fe] and [O/Fe] of APOGEE DR17, and the red dashed line denotes C/O = 0.5.\label{APOGEE_CO}}
\end{figure}

\begin{figure}[ht!] 
    \centering
    \includegraphics[width=0.48\textwidth]{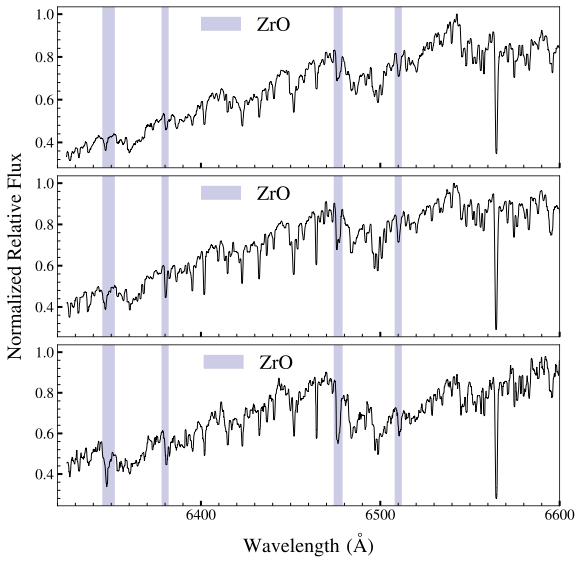}
    \caption{Three example spectra of the 18 S-type stars with C/O $<$ 0.5, which were estimated by [C/Fe] and [O/Fe] of APOGEE DR17.\label{Spectrum_CO_half}}
\end{figure}

In addition, we cross-matched the 2939 stars with APOGEE DR17, obtained [C/Fe] and [O/Fe] of 132 S-type stars, and calculated their C/Os using the method provided by \cite{2016ApJ...831...20B}. Figure \ref{APOGEE_CO} shows the C/O distribution of the 132 S-type stars, most of them have C/O larger than 0.5, and there are only 18 stars with C/O less than 0.5. We manually checked the spectra of these 18 S-type stars, which also have obvious ZrO bands as in Fig.\ref{Spectrum_CO_half}, and still retain them in the 2939 S-type star sample. The C/O distributions in Fig.\ref{MARCS_CO} and Fig.\ref{APOGEE_CO} are significantly different, most C/Os in Fig.\ref{MARCS_CO} are  generally higher (between 0.9 and 1.0) than the average expected C/O values, while those in Fig.\ref{APOGEE_CO} are closer to the expectations of values. For individual stars, the C/O discrepancy is obvious. As in the case of J050833.95+491212.8, the C/O values from APOGEE and MARCS models are 0.51 and 0.95, respectively, which are differ significantly. This disparity highlights that deriving C/O ratios from the $(V - K, J - K)$ diagram may be a less reliable method.

We cross-matched 2939 S-type stars with SIMBAD, whose information was derived entirely from the literature, and the detailed parameters of 1553 common stars are shown in Table \ref{tab:SIMBAD_inf}. The first, second, and third columns list the main type, other types, and the number of each main type, respectively. From this table, 94 stars have already been identified as S-type stars and/or candidates in previous literatures; thus, 2306 among the 2939 stars (excluding the S-type stars also in Paper I ) were reported as S-type stars for the first time, making it the largest catalog of S-type stars to date, as presented in Table \ref{tab:catalog}. The catalog can be available from the link: \url{https://paperdata.china-vo.org/chenjing/Sstar\_LAMOSTDR10.zip}.

Through the $M_{K} ~vs.~ W_{\mathrm{RP, BP-RP}} - W_{K,  J-K}$ diagram, most of 2939 stars are indeed belong to the O-rich AGB phase. From C/Os obtained by the two methods, they are all in the range of 0.5-1.0, and the C/O distribution from APOGEE is more consistent with the expectation. Furthermore, after cross-matching with SIMBAD, 91 S-type stars and 3 possible S-type stars have been reported in literature \citep{ 1975AbaOB..47....3D, 1979A&AS...38..335M, 1980ApJS...43..379K, 1984PW&SO...3....1S,}. To sum up, these analyses support the result that the 2939 stars are S-type stars.

\begin{deluxetable}{lcc}
    \tabletypesize{\scriptsize}
    \setlength\tabcolsep{2pt}
    \tablewidth{1pt} 
    \tablecaption{Information from SIMBAD for the 1553 stars.\label{tab:SIMBAD_inf}}
    \tablehead{$\mathrm{Main Type}$ & $\mathrm{Other Types}$ & Number}
    \startdata
    LPV* \tnote{a} & * $\mid$ LP* $\mid$ V* $\mid$ IR & 589 \\ 
	Star \tnote{b}& * $\mid$ IR & 561  \\
	V* \tnote{c} & * $\mid$ IR* & 183 \\
	S* \tnote{d} & LP* $\mid$ V* $\mid$ IR $\mid$ * & 91 \\
	Mira \tnote{e} & LP* $\mid$ V* $\mid$ IR $\mid$ * & 52 \\
	$\mathrm{LP^*\_Candidate}$ \tnote{f} & * $\mid$ IR $\mid$ V* & 36 \\
        SB* \tnote{g} &  * $\mid$ IR & 16  \\
	$\mathrm{S^*\_Candidate}$ \tnote{h} & S*? $\mid$ IR $\mid$ * $\mid$ S* & 3 \\
        C* \tnote{i} & * $\mid$ C*? $\mid$ IR $\mid$ LP*? $\mid$ V* & 3 \\
        RGB* \tnote{j} & * $\mid$ IR $\mid$ PM*? $\mid$ RG* & 3 \\
        IR \tnote{k} & * $\mid$ IR & 3  \\
        Em* \tnote{l} & * $\mid$ EM* $\mid$ IR* & 2 \\
        PulsV* \tnote{m} & * $\mid$ IR* $\mid$ LP* $\mid$ Pu* $\mid$ V* & 2 \\
        RotV* \tnote{n} & * $\mid$ IR* $\mid$ Ro* & 2 \\
        $\mathrm{EB^*\_Candidate}$ \tnote{o} & * $\mid$ EB?$\mid$ IR* & 2 \\
	AGB* \tnote{p} & LP* $\mid$ V* $\mid$ IR $\mid$ * & 1 \\
	NIR \tnote{q} & * $\mid$ IR & 1  \\ 
	$\mathrm{C^*\_Candidate}$ \tnote{r} & C*? $\mid$ IR $\mid$ *  & 1 \\
	$\mathrm{Mi^*\_Candidate}$ \tnote{s} & * $\mid$ IR $\mid$ Mi? $\mid$ V* & 1 \\
	$\mathrm{Orion\_V^*}$ \tnote{t} & Or* $\mid$ V* $\mid$ IR & 1  \\
    \enddata
    \tablecomments{
            $^a$ Long-period variable (LPV) star. \\
            $^b$ Star. \\
            $^c$ Variable star.\\
            $^d$ S-type star.\\
            $^e$ Variable star of Mira Cet type.\\
            $^f$ Long-period variable candidate.\\
            $^g$ Spectroscopic binary (SB).\\
            $^h$ Possible S-type star.\\
		$^i$ Carbon star.\\
		$^j$ Red giant branch (RGB) star.\\
		$^k$ Infra-Red source.\\
		$^l$ Emission-line star.\\
            $^m$ Pulsating variable star.\\
            $^n$ Rotationally variable star.\\
            $^o$ Eclipsing binary candidate.\\
            $^p$ Asymptotic giant branch star.\\
            $^q$ Near-infrared (NIR) source.\\
            $^r$ Possible carbon star.\\
            $^s$ Mira candidate.\\
            $^t$ Variable star of Orion type.}
\end{deluxetable}

\begin{deluxetable*}{cccccccccccccccccc}
    \tabletypesize{\scriptsize}
    \setlength\tabcolsep{2pt}
    \tablewidth{1pt} 
    \tablecaption{Ten Examples of the 2939 S-type Stars in This Work.\label{tab:catalog}}
    \tablehead{$\mathrm{Designation^{1}}$ & $\mathrm{R. A.}$ & $\mathrm{Decl.}$ & $J^2$ & $H^3$ & $K^4$ & $\mathrm{[09]^5}$ & $\mathrm{[18]^6}$ & $\mathrm{[12]^7}$ & $\mathrm{[25]^8}$ & $\mathrm{type\_SIMBAD^9}$ & $\mathrm{In Paper I^{10}}$ & \dcolhead{\mathrm{C/O_M^{11}}} & \dcolhead{\mathrm{C/O_A^{12}}} & \dcolhead{\mathrm{Class_X^{13}}} & \dcolhead{\mathrm{Class_C^{14}}} & \dcolhead{\mathrm{Class_S^{15}}} & \dcolhead{\mathrm{E_{RV}^{16}}} \\
     &\colhead{(degree)}& \colhead{(degree)}& \colhead{(mag)}& \colhead{(mag)}& \colhead{(mag)}& \colhead{(Jy)} & \colhead{(Jy)}& \colhead{(Jy)}& \colhead{(Jy)}&&&&&&&&}
    \startdata
    J021144.49+490442.6 & 32.9353774 & 49.0785078 & 6.03 & 5.12 & 4.80 & 0.76 & 0.18 & 0.62 & 0.15 & LPV* & N & 0.5 & ~ & E & E & E & F\\
    J022802.75+594610.0 & 37.0114850 & 59.7694670 & 4.39 & 3.48 & 3.10 & 3.12 & 0.79 & 2.30 & 0.51 & V* & N & 0.75 & ~ & E & E & E & F \\ 
    J024159.41+435716.7 & 40.4975440 & 43.9546410 & 6.36 & 5.42 & 5.10 & 0.60 & 0.13 & 0.48 & 0.28 & Star & N & 0.75 & ~ & E & E & E & F \\
    J024201.73+690006.3 & 40.5072100 & 69.0017600 & 6.90 & 5.71 & 5.29 & 0.54 & 0.13 & 0.38 & 0.26 & Star & N & 0.97 & ~ & E & E & E & F\\ 
    J025629.73+623634.5 & 44.1238890 & 62.6095960 & 3.44 & 2.54 & 2.13 & 7.70 & 1.83 & 5.59 & 1.45 & V* & N & 0.99 & ~ & E & E & E & F\\ 
    J061147.90+221021.6 & 92.9495937 & 22.1726712 & 6.09 & 4.92 & 4.52 & 1.37 & 0.54 & 1.26 & 0.50 & LPV* & Y & 0.99 & ~ & I & I & I & T\\ 
    J061207.47-063700.2 & 93.0311400 & -6.61672300 & 8.50 & 7.43 & 7.07 & 0.34 & 0.16 & 0.30 & 0.16 & LPV* & Y & 0.95 & 0.82 & I & I & I & F\\ 
    J061345.33+113139.5 & 93.4388960 & 11.5276490 & 6.86 & 5.74 & 5.28 & 0.68 & 0.20 & 0.46 & 0.29 & LPV* & N & 0.95 & ~ & I & I & I & T\\ 
    J061415.86+225258.9 & 93.5661190 & 22.8830500 & 7.74 & 6.45 & 5.96 & 0.44 & 0.24 & 0.33 & 0.55 & LPV* & N & 0.95 & ~ & I & I & I & F\\ 
    J061549.22+250040.5 & 93.9551240 & 25.0112540 & 6.61 & 5.46 & 5.05 & 1.21 & 0.48 & 1.34 & 0.55 & S* & Y & 0.95 & ~ & I & I & I & F\\
    \enddata
    \tablecomments{Column 4 to 10 are the magnitudes and flux from 2MASS, AKARI, and IRAS. Columns 13 and 14 are C/O estimated by the MARCS models of S-type stars and [C/Fe] and [O/Fe] from APOGEE DR17. The last four columns denote the intrinsic and extrinsic classification results by four methods, which will be introduced in section \ref{sec:distinguish} and section \ref{sec:discuss}.
    (1) The designation from LAMOST DR10. 
    (2) - (4) The J, H, and K magnitudes from 2MASS. 
    (5) and (6) The flux of [09] and [18] from AKARI. 
    (7) and (8) The flux of [12] and [25] from IRAS. 
    (9) The main type from the SIMBAD. 
    (10) Whether it's in Paper I.
    (11) and (12) The C/O estimated by the MARCS S-type star model and the [C/Fe] and [O/Fe] from APOGEE. 
    (13) - (15) The classified result from XGBoost model, color-color diagram, and optical spectrum, respectively.
    (16) Sources with large variations in RVs.
    }
\end{deluxetable*}

\section{Distinguish Intrinsic and Extrinsic S-type Stars Using Infrared Photometric Data} \label{sec:distinguish}

In this section, the XGBoost model and the color-color diagram were studied to classify intrinsic and extrinsic stars with the IR photometric data from 2MASS, WISE, the infrared astronomical mission AKARI, and IRAS, and the two methods were adopted to classify the 2939 S-type stars.

\subsection{The Machine-learning method}

To investigate which IR photometric features are most important for distinguishing between intrinsic and extrinsic S-type stars, we adopted the XGBoost algorithm in this work. Recently, XGBoost has been widely applied to astronomy and showed its capabilities in dealing with astronomical problems, such as, distinguishing M giants from M dwarfs for spectral surveys \citep{2019ApJ...887..241Y}, and selecting quasar candidates with photometric data \citep{2019MNRAS.485.4539J}.

\subsubsection{Data}\label{sec:sample}

In order to train and test the XGBoost Classifier, it is necessary to collect as many correctly classified intrinsic and extrinsic S-type stars as possible. \cite{2019AJ....158...22C} found 151 extrinsic S-type stars and 190 intrinsic S-type stars. \cite{2018AA...620A.148S, 2019AA...625L...1S, 2020AA...635L...6S, 2021AA...650A.118S} used the presence of Tc lines in the HERMES high-resolution spectrum to classify some intrinsic and extrinsic S-type stars. Due to the presence or absence of the Tc line as the most direct way to distinguish between intrinsic and extrinsic S-type stars, according to \cite{2018AA...620A.148S, 2019AA...625L...1S, 2020AA...635L...6S, 2021AA...650A.118S}, we have corrected the classification results of a fraction of S-type stars in \cite{2019AJ....158...22C}. Table \ref{tab:oldsample} lists several modified S-type stars, of which 7 were changed from extrinsic to intrinsic, and 3 were changed from intrinsic to extrinsic. Table \ref{tab:newsample} lists intrinsic and extrinsic S-type stars not in \cite{2019AJ....158...22C} but in other literature \citep{2018AA...620A.148S, 2019AA...625L...1S,  2019AA...626A.127J, 2021AA...650A.118S}. Eventually, we collected 151 extrinsic and 205 intrinsic S-type stars as positive samples of XGBosst algorithm in this work. In Tables \ref{tab:oldsample} and \ref{tab:newsample}, the third column shows the number in the General Catalog of Galactic S stars from \cite{1984PW&SO...3....1S}, which is the largest catalog of S-type stars before.

\begin{deluxetable*}{c|ccccc}[ht!]
\setlength\tabcolsep{10pt}
\caption{Corrected the Classification Results of Intrinsic and Extrinsic Stars in 
\cite{2019AJ....158...22C}.}
\tablehead{
\colhead{Class} & \colhead{Name} & \dcolhead{\mathrm{CSS^c}} & \colhead{R.A. (h:m:s)} & \colhead{Decl. (d:m:s)} & \dcolhead{\mathrm{Reference^d}}
}
\startdata
\multirow{7}{*}{$\rm E \Rightarrow I^a $} & $\rm BD+79^{\circ}~156$ & 106 & 04 55 44.9993026824 & +79 59 59.984445300 & \cite{2020AA...635L...6S} \\
 & HD 288833 & 233 & 06 36 12.1133753592 & +01 57 29.168329464 & \cite{2021AA...650A.118S} \\
 & $\rm V^{\ast}~KR~CMa$ & 265 & 06 48 50.5905123984 & -20 25 32.256147180 & \cite{2021AA...650A.118S} \\
 & $\rm BD+34^{\circ}~1698$ & 413 & 07 52 53.2125114216 & +34 36 50.804714952 & \cite{2019AA...625L...1S} \\
 & CSS 454 & 454 & 08 04 40.8861230688 & -03 04 25.762616760 & \cite{2021AA...650A.118S} \\
 & HD 357941 & 1190 & 20 07 43.9040448432 & -01 36 10.173925152 & \cite{2019AA...625L...1S} \\
 & $\rm V^{\ast}~SX~Peg$ & 1309 & 22 50 24.8246941896 & +17 53 36.514618152 & \cite{2021AA...650A.118S} \\ \hline
 \multirow{3}{*}{$\rm I \Rightarrow E^{b}$} & $\rm V^{\ast}~FX~CMa$ & 350 & 07 27 03.9357410184 & -11 43 14.442010248 & \multirow{3}{*}{\cite{2018AA...620A.148S}} \\
 & HD 63733 & 411 & 07 49 45.7502346456 & -19 00 24.859707408 & \\
 & V530 Lyr & 1053 & 18 30 34.5709214544 & +36 14 56.999721696 & \\
\enddata
\begin{tablenotes}
\footnotesize
\item $^a$ Change the classification result from extrinsic to intrinsic.
\item $^b$ Change the classification result from intrinsic to extrinsic.
\item $^c$ The identifier in the General Catalog of Galactic S stars, which is the largest catalog of S-type stars before.
\item $^d$ The latest literature of studying intrinsic and extrinsic S-type stars.
\end{tablenotes}
\end{deluxetable*} \label{tab:oldsample}

\begin{deluxetable*}{c|ccccc}
\setlength\tabcolsep{10pt}
\caption{Intrinsic and Extrinsic S-type Stars Selected from Literature Other Than \cite{2019AJ....158...22C}.}
\label{tab:newsample}
\tablehead{
\colhead{Class} & \colhead{Name} & \dcolhead{\mathrm{CSS^a}} & \colhead{R.A. (h:m:s)} & \colhead{Decl. (d:m:s)} & \dcolhead{\mathrm{Reference^b}}
}
\startdata
\multirow{4}{*}{E} & AB Col & 174 & 05 55 34.5239404104 & -28 57 13.295878812 & \multirow{4}{*}{\cite{2018AA...620A.148S}} \\
 & TYC 5971-534-1 & 302 & 07 03 55.4329806864 & -19 06 28.546209324 & \\
 & $\rm BD -22^{\circ}~1742$ & 318 & 07 11 26.8432250856 & -22 24 04.333260168 &  \\
 & $\rm BD +69^{\circ}~524$ & 612 & 09 35 36.4050766008 & +69 09 25.192466352 &  \\ \hline
\multirow{11}{*}{I}& $\rm S1^{\ast}~111$ & 151 & 05 40 22.9411904424 & +10 44 04.947970848 & \cite{2021AA...650A.118S} \\
 & IRAS 05387+0137 & 154 & 05 41 23.6747573400 & +01 38 34.704935304 & \cite{2019AA...625L...1S} \\
 & IRAS 06000+1023& 182 & 06 02 52.2202725744 & +10 22 58.805334444 & \cite{2019AA...625L...1S} \\
  & $\rm BD -18^{\circ}~2608$ & 597 & 09 12 58.2093870072 & -18 43 31.784979360 & \cite{2021AA...650A.118S} \\
 & UY Cen & 816 & 13 16 31.8299380416 & -44 42 15.757651800 & \cite{2018AA...620A.148S} \\
 & R Cam & 856 & 14 17 51.0361093608 & +83 49 53.840253360 & \cite{2021AA...650A.118S} \\
 & V812 Oph & 997 & 17 41 31.9391647512 & +06 43 41.330915364 & \cite{2021AA...650A.118S} \\
 & W Aql & 1115 & 19 15 23.3572741560 & -07 02 50.333886492 & \cite{2021AA...650A.118S} \\
 & V4638 Sgr & 1140 & 19 34 00.3908425752 & -20 55 56.148163428 & \cite{2019AA...626A.127J} \\
 & Vy 12 & 1152 & 19 37 22.8503043720 & +67 11 19.999732020 & \cite{2021AA...650A.118S} \\
 & $\rm BD +31^{\circ}~4391$ & 1267 & 21 15 41.1687655320 & +31 45 39.469496148 & \cite{2019AA...626A.127J} \\
\enddata
\begin{tablenotes}
\footnotesize
\item $^a$ The identifier in the General Catalog of Galactic S stars, which is the largest catalog of S-type stars before.
\item $^b$ The origin literature of intrinsic and extrinsic S-type stars.
\end{tablenotes}
\end{deluxetable*} 

In this work, we selected the photometric data of 2MASS, AllWISE, AKARI, and IRAS surveys to study which IR magnitudes or colors are the most important for classifying intrinsic and extrinsic S-type stars. A detailed description of the four surveys is given below:
\begin{itemize}
    \item 2MASS\citep{2006AJ....131.1163S} has constructed a near-IR $J-$ (1.25$\mu$m), $H-$ (1.65 $\mu$m), and $Ks$-band (2.16 $\mu$m) image of the entire sky, and the 2MASS Point Source Catalog contains photometry and astrometry for 471 million objects.
    \item WISE \citep{2010AJ....140.1868W} is a mid-IR survey of the entire sky, offering four photometric bands of $W_1$, $W_2$, $W_3$, and $W_4$ with wavelengths centered at 3.4, 4.6, 12, and 22 $\mu$m, respectively. The AllWISE source catalog, which was released in 2013, provides positions and four-band fluxes for over 747 million objects. Due to the interstellar dust emitting wavelengths that contaminate $W_3$ and $W_4$ photometry, we only used $W_1$ and $W_2$ magnitudes in this work \citep{2022A&A...662A.125F}.
    \item IRAS conducted an all-sky survey \citep{1988iras....1.....B}; the point-source catalog (PSC) and the faint source catalog (FSC) provided photometric data at four bands (12, 25, 60, 100 $\mu$m). \cite{2015A&C....10...99A} cross-correlated these two catalogs and created the IRAS PSC/FSC Combined Catalogue, which contains 345,162 sources. 
    \item AKARI \citep{2007PASJ...59S.369M} was designed as an All-Sky Survey mission in the IR region, and covers more than 90\% of the whole sky with a higher spatial resolution and a wider wavelength coverage than that of the previous IRAS all-sky survey. AKARI provided the PSC data at two bands (9 and 18 $\mu$m; \cite{2010A&A...514A...1I}) for 870,973 sources.
\end{itemize}

The collected 151 extrinsic and 205 intrinsic S-type stars were cross-matched with 2MASS, ALLWISE, AKARI, and IRAS, within $5^{''}$, respectively. Table \ref{tab:match} lists the numbers of the common stars of the extrinsic and intrinsic stars with the four surveys, and these common stars were used for training and testing the XGBoost model. The intrinsic S-type stars were used as the positive sample, the extrinsic stars were treated as the negative sample, and the positive and negative samples were randomly divided into the training and testing samples according the ratio of 8:2.

\begin{deluxetable}{cccccc}
\setlength\tabcolsep{2pt}
\caption{The Number of Common Stars of Known S-type Stars Collected in Section \ref{sec:sample} after Cross-matching with Four Infrared Surveys.}
\tablehead{
\colhead{Class} & \colhead{Number} & \colhead{2MASS} & \colhead{AllWISE} & \colhead{AKARI} & \colhead{IRAS}
}
\startdata
intrinsic S-type stars & 205 & 204 & 198 & 200 & 198 \\
extrinsic S-type stars & 151 & 151 & 147 & 149 & 143 \\
\enddata
\end{deluxetable} \label{tab:match}

\subsubsection{Input features} \label{Input_color}

Table \ref{tab:IRband} lists the IR bands used in this work. For each band, the reference wavelength ($\rm \lambda_{ref}$) and zero-magnitude flux (ZMF) value are also shown. The color index is defined by 
\begin{gather}
    \mathrm{m_{\lambda_1} - m_{\lambda_2} = -2.5log_{10} \frac{F_{\lambda_1} / ZMF_{\lambda_1}}{F_{\lambda_2} / ZMF_{\lambda_2}}},
\end{gather}
where $\rm ZMF_{\lambda i}$ is the ZMF at given wavelength. The calculation of the color index except $J$, $H$, $K$, W1, and W2 is calculated directly by the difference in magnitude, and the other [09], [12], [18], [25] need to convert the flux to magnitude first, and then calculate the color index. In addition, due to intrinsic S-type stars that are in the TP-AGB stage and due to most known extrinsic S-type stars that are assumed to be members of the RGB \citep{2011ApJS..196....5O}, they should also be distinguished in the color-magnitude diagrams. So the absolute magnitude of each band was also treated as input features to train the models, and the absolute magnitudes can be calculated by
\begin{gather}
    M = m + 5 - 5log_{10}r,
\end{gather}
where $m$ is apparent magnitude, and $r$ is heliocentric distance. It is naive to directly use the inverse of Gaia parallax to estimate heliocentric distance; thus, we adopted the Monte Carlo's method as in \cite{2022MNRAS.515..767M} to estimate the distance.

\begin{deluxetable}{ccccc}
\caption{The Zero-magnitude Flux Values for Magnitudes of Four IR Surveys.}
\tablehead{
\colhead{Telescope} & \colhead{Band} & \dcolhead{\rm \lambda_{ref} (\mu m)} & \colhead{ZMF (Jy)} & Reference
}
\startdata
\multirow{3}{*}{2MASS} & $J$ & 1.235 & 1594 & \multirow{3}{*}{\cite{2003AJ....126.1090C}} \\
 & $H$ & 1.662 & 1024 & \\
 & $K$ & 2.159 & 666.7 & \\ \hline
\multirow{2}{*}{AllWISE} & W1 & 3.4 & 306.682 & \multirow{2}{*}{\cite{2011ApJ...735..112J}} \\
 & W2 & 4.6 & 170.663 & \\ \hline
\multirow{2}{*}{AKARI} & [9] & 9 & 56.262 & \multirow{2}{*}{\cite{2007PASJ...59S.369M}} \\
 & [18] & 18 & 12.001 & \\ \hline
\multirow{2}{*}{IRAS} & [12] & 12 & 28.3 & \multirow{2}{*}{\cite{1988iras....1.....B}} \\
 & [25] & 25 & 6.73 & \\
\enddata
\end{deluxetable} \label{tab:IRband}

For the input colors of the four surveys, the galactic extinction correction is needed to be considered. 
The extinctions of $J$, $H$, $K$ bands were estimated by the $\rm dustmaps$ Python package as in section \ref{sec:identify}. The Python package $dust\_extinction$\footnote{\url{https://github.com/karllark/dust\_extinction}} provides a suite of models of interstellar dust extinction curves, and we chose the extinction curve of \cite{2006ApJ...637..774C} to estimate the extinction of W1, W2, [09], [18], [12], and [25] bands.

In this work, we used 15 input features to test which of them has the best accuracy to distinct intrinsic and extrinsic S-type stars when using XGBoost algorithm, and they were abbreviated as 2MASS, AllWISE, AKARI, IRAS, 2MASS + ALLWISE (2W), 2MASS + AKARI (2A), 2MASS + IRAS (2I), ALLWISE + AKARI (WA), ALLWISE + IRAS (WI), AKARI + IRAS (AI), 2MASS + ALLWISE + AKARI (2WA), 2MASS + ALLWISE + IRAS (2WI), 2MASS + AKARI + IRAS (2AI), ALLWISE + AKARI + IRAS (WAI), and 2MASS + ALLWISE + AKARI + IRAS (2WAI), respectively. 2MASS, AllWISE, AKARI, and IRAS represent input features, which only use absolute magnitudes and colors from one of 2MASS, AllWISE, AKARI, and IRAS surveys; 2W, 2A, 2I, WA, WI, and AI are input features that use magnitudes and colors of two surveys; 2WA, 2WI, 2AI, and WAI denote input features that use three surveys; and 2WAI indicates the input feature that uses all four surveys. The magnitudes and colors used by each input feature, which were marked by the `` $\surd$, " are listed in Table \ref{tab:inputcolors}.

\begin{deluxetable*}{cccccccccccccccc}
    \tabletypesize{\scriptsize}
    \setlength\tabcolsep{3pt}
    \tablewidth{1pt} 
    \tablecaption{Absolute Magnitudes and Colors Used by 15 Input Features When Using XGBoost Model \label{tab:inputcolors}}
    \tablehead{Absolute Magnitude and/or Color & 2MASS & AllWISE & AKARI & IRAS & 2W & 2A & 2I & WA & WI & AI & 2WA & 2WI & 2AI & WAI & 2WAI}
    \colnumbers
    \startdata
    $\mathrm{M_J}$ & $\surd$ & & & & $\surd$ & $\surd$ & $\surd$ & & & & $\surd$ & $\surd$ & $\surd$ & & $\surd$ \\
    $\mathrm{M_H}$ & $\surd$ & & & & $\surd$ & $\surd$ & $\surd$ & & & & $\surd$ & $\surd$ & $\surd$ & & $\surd$ \\
    $\mathrm{M_K}$ & $\surd$ & & & & $\surd$ & $\surd$ & $\surd$ & & & & $\surd$ & $\surd$ & $\surd$ & & $\surd$ \\
    $\mathrm{M_{W1}}$ & & $\surd$ & & & $\surd$ & & & $\surd$ & $\surd$ & & $\surd$ & $\surd$ & & $\surd$ & $\surd$ \\
    $\mathrm{M_{W2}}$ & & $\surd$ & & & $\surd$ & & & $\surd$ & $\surd$ & & $\surd$ & $\surd$ & & $\surd$ & $\surd$ \\
    $\mathrm{M_{[09]}}$ & & & $\surd$ & & & $\surd$ & & $\surd$ & &$\surd$ & $\surd$ & & $\surd$ & $\surd$ & $\surd$ \\
    $\mathrm{M_{[12]}}$ & & & & $\surd$ & & & $\surd$ & & $\surd$ & $\surd$ & & $\surd$ & $\surd$ & $\surd$ & $\surd$ \\
    $\mathrm{M_{[18]}}$ & & & $\surd$ & & & $\surd$ & & $\surd$ & & $\surd$ & $\surd$ & & $\surd$ & $\surd$ & $\surd$ \\
    $\mathrm{M_{[25]}}$ & & & & $\surd$ & & & $\surd$ & & $\surd$ & $\surd$ & & $\surd$ & $\surd$ & $\surd$ & $\surd$ \\
    $\mathrm{(J-H)_0}$ & $\surd$ & & & & $\surd$ & $\surd$ & $\surd$ & & & & $\surd$ & $\surd$ & $\surd$ & & $\surd$ \\
    $\mathrm{(J-K)_0}$ & $\surd$ & & & & $\surd$ & $\surd$ & $\surd$ & & & & $\surd$ & $\surd$ & $\surd$ & & $\surd$ \\
    $\mathrm{(J-W_1)_0}$ & & & & & $\surd$ & & & & & & $\surd$ & $\surd$ & & & $\surd$ \\
    $\mathrm{(J-W_2)_0}$ & & & & & $\surd$ & & & & & & $\surd$ & $\surd$ & & & $\surd$ \\
    $\mathrm{(J-[09])_0}$ & & & & & & $\surd$ & & & & & $\surd$ & $\surd$ & & & $\surd$ \\
    $\mathrm{(J-[12])_0}$ & & & & & & & $\surd$ & & & & & $\surd$ & $\surd$ & & $\surd$ \\
    $\mathrm{(J-[18])_0}$ & & & & & & $\surd$ & & & & & $\surd$ & & $\surd$ & & $\surd$ \\
    $\mathrm{(J-[25])_0}$ & & & & & & & $\surd$ & & & & & $\surd$ & $\surd$ & & $\surd$ \\
    $\mathrm{(H-K)_0}$ & $\surd$ & & & & $\surd$ & $\surd$ & $\surd$ & & & & $\surd$ & $\surd$ & $\surd$ & & $\surd$ \\
    $\mathrm{(H-W_1)_0}$ & & & & & $\surd$ & & & & & & $\surd$ & $\surd$ & & & $\surd$ \\
    $\mathrm{(H-W_2)_0}$ & & & & & $\surd$ & & & & & & $\surd$ & $\surd$ & & & $\surd$ \\
    $\mathrm{(H-[09])_0}$ & & & & & & $\surd$ & & & & & $\surd$ & & $\surd$ & & $\surd$ \\
    $\mathrm{(H-[12])_0}$ & & & & & & & $\surd$ & & & & & $\surd$ & $\surd$ & & $\surd$ \\
    $\mathrm{(H-[18])_0}$ & & & & & & $\surd$ & & & & & $\surd$ & & $\surd$ & & $\surd$ \\
    $\mathrm{(H-[25])_0}$ & & & & & & & $\surd$ & & & & & $\surd$ & $\surd$ & & $\surd$ \\
    $\mathrm{(K-W_1)_0}$ & & & & & $\surd$ & & & & & & $\surd$ & $\surd$ & & & $\surd$ \\
    $\mathrm{(K-W_2)_0}$ & & & & & $\surd$ & & & & & & $\surd$ & $\surd$ & & & $\surd$ \\
    $\mathrm{(K-[09])_0}$ & & & & & & $\surd$ & & & & & $\surd$ & & $\surd$ & & $\surd$ \\
    $\mathrm{(K-[12])_0}$ & & & & & & & $\surd$ & & & & & $\surd$ & $\surd$ & & $\surd$ \\
    $\mathrm{(K-[18])_0}$ & & & & & & $\surd$ & & & & & $\surd$ & & $\surd$ & & $\surd$ \\
    $\mathrm{(K-[25])_0}$ & & & & & & & $\surd$ & & & & & $\surd$ & $\surd$ & & $\surd$ \\
    $\mathrm{(W_1-W_2)_0}$ & & $\surd$ & & & $\surd$ & & & $\surd$ & $\surd$ & & $\surd$ & $\surd$ & & $\surd$ & $\surd$ \\
    $\mathrm{(W_1-[09])_0}$ & & & & & & & & $\surd$ & & & $\surd$ & & & $\surd$ & $\surd$ \\
    $\mathrm{(W_1-[12])_0}$ & & & & & & & & & $\surd$ & & & $\surd$ & & $\surd$ & $\surd$ \\
    $\mathrm{(W_1-[18])_0}$ & & & & & & & & $\surd$ & & & $\surd$ & & & $\surd$ & $\surd$ \\
    $\mathrm{(W_1-[25])_0}$ & & & & & & & & & $\surd$ & & & $\surd$ & & $\surd$ & $\surd$ \\
    $\mathrm{(W_2-[09])_0}$ & & & & & & & & $\surd$ & & & $\surd$ & & & $\surd$ & $\surd$ \\
    $\mathrm{(W_2-[12])_0}$ & & & & & & & & & $\surd$ & & & $\surd$ & & $\surd$ & $\surd$ \\
    $\mathrm{(W_2-[18])_0}$ & & & & & & & & $\surd$ & & & $\surd$ & & & $\surd$ & $\surd$ \\
    $\mathrm{(W_2-[25])_0}$ & & & & & & & & & $\surd$ & & & $\surd$ & & $\surd$ & $\surd$ \\
    $\mathrm{([09]-[12])_0}$ & & & & & & & & & & $\surd$ & & & $\surd$ & $\surd$ & $\surd$ \\
    $\mathrm{([09]-[18])_0}$ & & & $\surd$ & & & $\surd$ & & $\surd$ & & $\surd$ & $\surd$ & & $\surd$ & $\surd$ & $\surd$ \\
    $\mathrm{([09]-[25])_0}$ & & & & & & & & & &$\surd$ & & & $\surd$ & $\surd$ & $\surd$ \\
    $\mathrm{([12]-[18])_0}$ & & & & & & & & & & $\surd$ & & & $\surd$ & $\surd$ & $\surd$ \\
    $\mathrm{([12]-[25])_0}$ & & & & $\surd$ & & & $\surd$ & & $\surd$ & $\surd$ & & $\surd$ & $\surd$ & $\surd$ & $\surd$ \\
    $\mathrm{([18]-[25])_0}$ & & & & & & & & & & $\surd$ & & & $\surd$ & $\surd$ & $\surd$ \\
    \enddata
        \tablecomments{\\
        Column (1): absolute magnitudes and colors used by each input features. 
        Column (2) - (5): input feature that only uses absolute magnitudes and colors from one of 2MASS, AllWISE, AKARI, and IRAS surveys.
        Column (6) 2W: input feature that uses absolute magnitudes and colors of 2MASS and AllWISE surveys.
        Column (7) 2A: input feature that uses absolute magnitudes and colors of 2MASS and AKARI surveys.
        Column (8) 2I: input feature that uses absolute magnitudes and colors of 2MASS and IRAS surveys.
        Column (9) WA: input feature that uses absolute magnitudes and colors of AllWISE and AKARI surveys.
        Column (10) WI: input feature that uses absolute magnitudes and colors of AllWISE and IRAS surveys.
        Column (11) AI: input feature that uses absolute magnitudes and colors of AKARI and IRAS surveys.
        Column (12) 2WA: input feature that uses absolute magnitudes and colors of 2MASS, AllWISE, and AKARI surveys.
        Column (13) 2WI: input feature that uses absolute magnitudes and colors of 2MASS, AllWISE, and IRAS surveys.
        Column (14) 2AI: input feature that uses absolute magnitudes and colors of 2MASS, AKARI, and IRAS surveys.
        Column (15) WAI: input feature that uses absolute magnitudes and colors of AllWISE, AKARI, and IRAS surveys.
        Column (16) 2WAI: input feature that uses absolute magnitudes and colors of 2MASS, AllWISE, AKARI, and IRAS surveys.}
\end{deluxetable*}

\subsubsection{Train XGBoost Model with the 15 Input Features}
\label{XGBoost_infraredcolor_results}

\begin{figure*}[ht!] 
    \centering
    \includegraphics[width=1.0\textwidth]{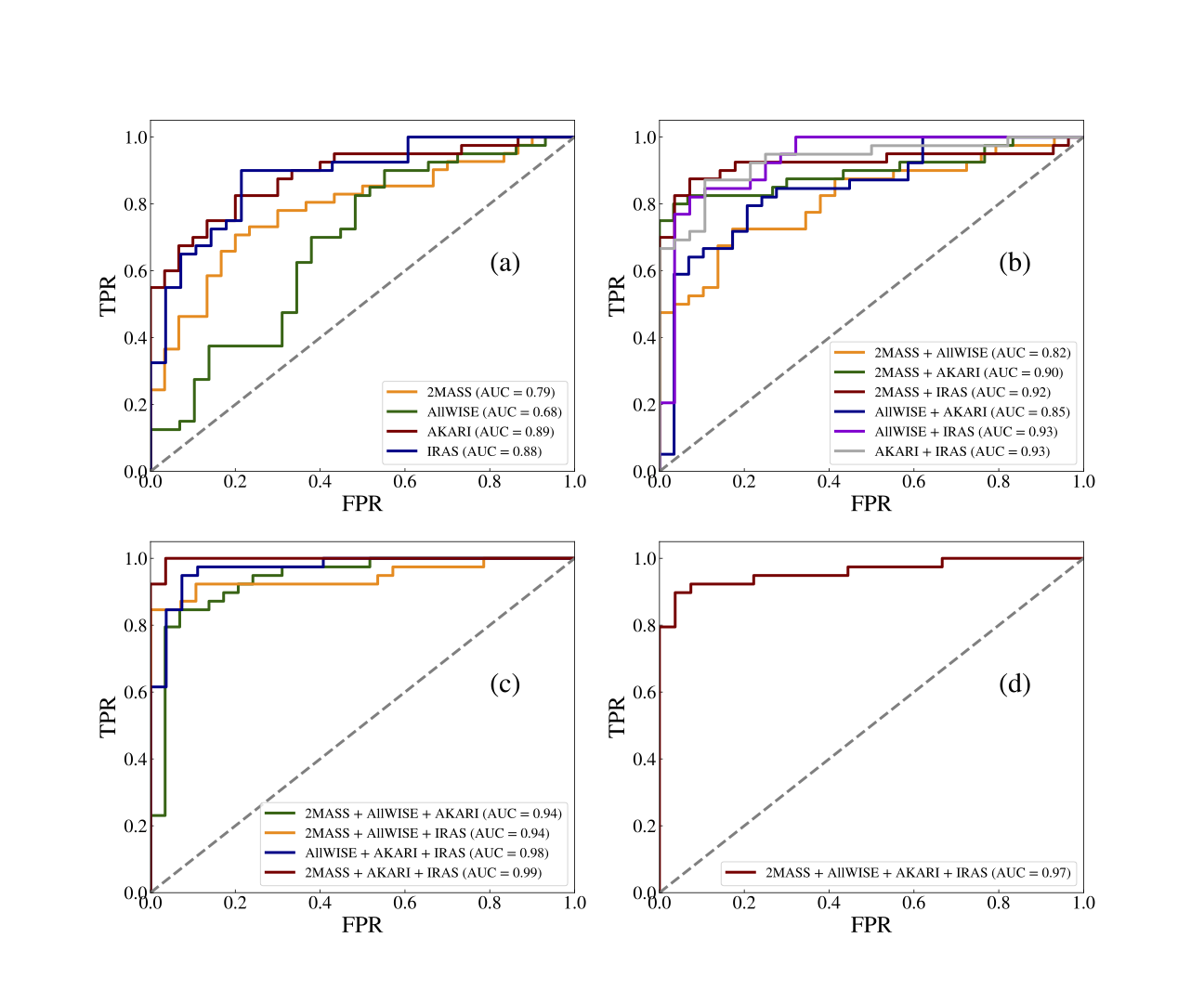}
    \caption{ROC curves for the 15 input features mentioned in subsection \ref{Input_color}. \label{ROC_curve}}
\end{figure*}

Using the 15 input features of the training and testing samples, we trained and tested the XGBoost model, respectively, and the classification results of the 15 input features were compared using the receiver operating characteristic (ROC) curves \citep{fawcett2006introduction}, which were used to characterize the binary classification ability of the XGBoost model, and  were shown in Fig.\ref{ROC_curve}.

The ROC curve represents the estimated true positive rate (TPR; the ratio between true positive and total positive cases) versus the estimated false positive rate (FPR; the ratio between false positive and total negative cases) for various probability thresholds. The TPR and FPR are defined as follows:
\begin{gather}
    \mathrm{TPR = \frac{TP}{TP+FN}},
    \mathrm{FPR = \frac{FP}{FP+TN}},
\end{gather}
where TP, FN, FP, and TN are same as formula \ref{1}. The better the classifier, the closer to the left y-axis and upper x-axis, i.e. it should maximize TPR, and minimize FPR values.

ROC curves were calculated when the 15 input features mentioned in subsection \ref{Input_color} were used separately, which were shown in Figure \ref{ROC_curve}. The area under each curve (AUC) that can be used as a tracer for model quality, and the larger value of AUC corresponds to higher accuracy of the classifier. From Fig.\ref{ROC_curve}, we can see that the accuracy of the classifier trained with only one survey feature (Fig.\ref{ROC_curve} (a)) is significantly lower than two (Fig.\ref{ROC_curve} (b)) or three more surveys (Fig.\ref{ROC_curve} (c) and Fig.\ref{ROC_curve} (d)) features. Since the XGBoost model trained with three or four IR survey data is more accurate, we recommend collecting as much IR survey data as possible when classifying intrinsic and extrinsic S-type stars, especially the characteristics of IRAS. In addition, no matter which model, as long as the IRAS absolute magnitudes and colors were included, the model accuracy became higher. In Figure \ref{ROC_curve}, when 2AI was selected as the input feautres of XGBoost, it has the largest AUC value of 99$\%$ and the highest accuracy of 95.52\%; thus, it was selected as the final input feature in this work for distinguishing intrinsic and extrinsic S-type stars with IR photometric data. Compared to 2AI input feature, 2WAI with more photometric data has instead a little bit lower AUC value and accuracy, which indicates that the addition of AllWISE data does not improve the accuracy of the model. As shown in Fig.\ref{ROC_curve} (a), the accuracy of the model trained only by W1 and W2 bands of AllWISE is lower than that of other surveys, which means that spectra in W1 and W2 bands likely have less discrepancy for intrinsic and extrinsic S-type stars. We will collect as many infrared spectra as possible to demonstrate that the W1 and W2 spectra are indeed less sensitive in distinguishing intrinsic and extrinsic S-type stars, and further investigate which molecular bands or atomic lines in IR band are more sensitive to intrinsic and extrinsic S-type stars in future.

When 2AI input feature was used, the trained XGBoost model parameters were as follows: gradient boosting trees were chosen as the structure of our model (i.e., booster: $gbtree$) to solve the nonlinear problem; the number of trees is 100 (i.e.,  $n\_estimators$ = 100), the maximum depth of a tree and the L2 regularization parameter were set to 12 (i.e., $max\_depth$ = 12) and 2 (i.e., $lambda$ = 2), respectively, which help to reduce the complexity of the model and prevent overfitting.

\begin{figure}[ht!] 
	\centering
        \includegraphics[width=0.5\textwidth]{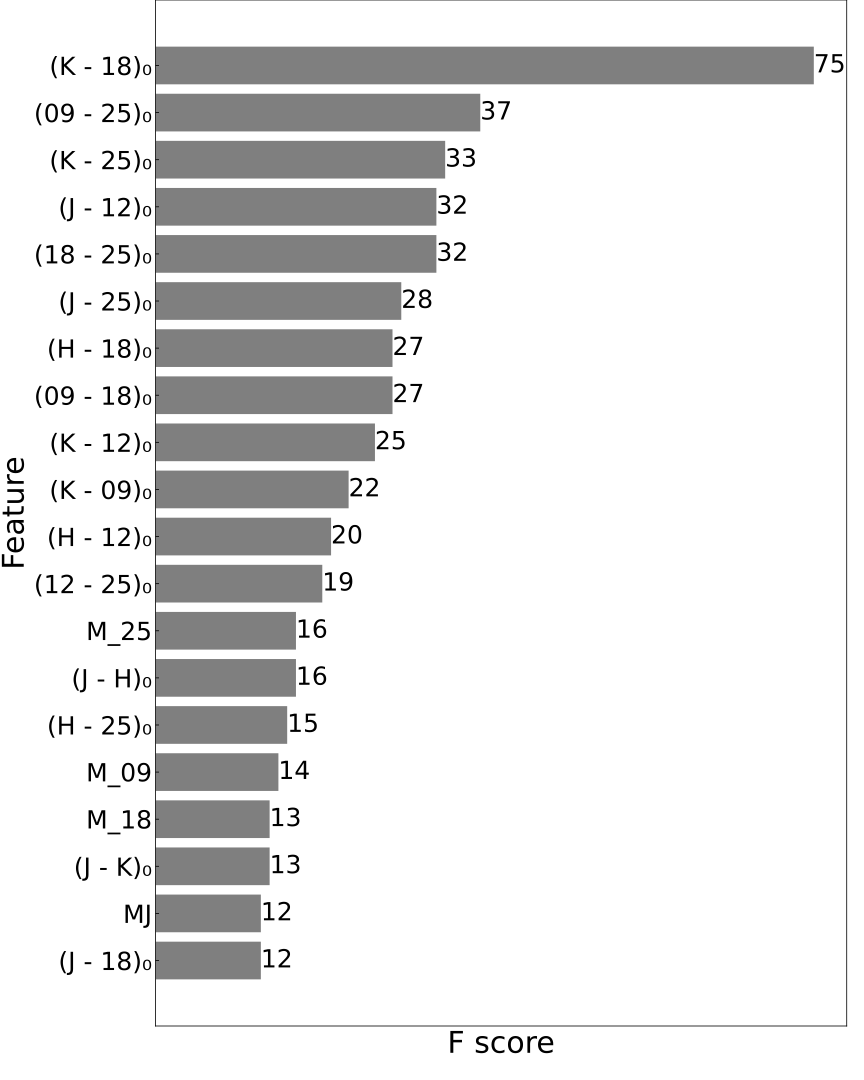}
	\caption{Input feature ranked based on their importance. \label{Feature_Importance}} 
\end{figure}

The default algorithm of native feature importance calculation within the XGBoost model was used to quantify the importance of each absolute magnitude or color used by the 2AI input feature, and the importance of each feature was estimated by calculating how much they are used to make key decisions. Specifically, the importance of a feature is a score that indicates how valuable it was when constructing the decision trees. Figure \ref{Feature_Importance} shows the top 20 features of the 2AI feature, they were ranked by the scores, and the first two most important colors of $(K - \mathrm{[18])_0}$ and $\mathrm{(09 - [25])_0}$ are thought to be due to the differences in silicate signatures between intrinsic and extrinsic S-type stars at 9.7 and 18$\mu$m, which requires further verification.

\subsection{The Color-Color Diagram Method}
\label{sec:color_color_method}

In the last subsection, XGBoost algorithm based on the tree model was adopted to distinguish intrinsic and extrinsic S-type stars with IR photometric data, and we also want to find more objective and accurate criteria to classify them with color-color diagram than in literature here. As we have known, the tree-based XGBoost model has the advantage of strong interpretability compared to the linear model; thus, we can get more objective IR color-color critria through the interpretability of tree-based XGBoost algorithm. To achieve this purpose, the XGBoost model with 2AI input feature trained in subsection \ref{XGBoost_infraredcolor_results} was used here, the weights of the decision trees were ignored, and each of the 100 decision trees in the model was used to repredict the test data. Finally, the decision tree with the highest accuracy was used to find the quantitative color-color criteria for distinguishing between intrinsic and extrinsic S-type stars.

\begin{figure*}[ht!] 
	\centering
        \includegraphics[width=1.0\textwidth]{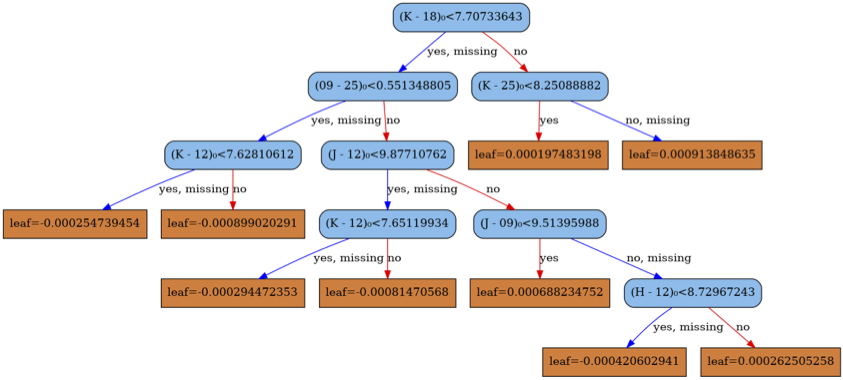}
	\caption{The most accurate decision tree of XGBoost model in this work. \label{Decision_Tree}} 
\end{figure*}

Figure \ref{Decision_Tree} shows the finally selected decision tree with the highest accuracy. The blue blocks represent the internal nodes, which mean the decision-making process; the yellow blocks represent the leaf nodes, which mean the decision result. The leaf node with leaf $>$ 0 denotes an S-type star was classified as intrinsic, and the node with leaf $<$ 0 denotes an S-type star was classified as extrinsic. The missing result in the tree represents that the colors have not been obtained from 2MASS, AKARI, and IRAS surveys, and XGBoost algorithm can get good predict results for such features with missing values.

\begin{figure*}[ht!] 
	\centering
        \includegraphics[width=1.0\textwidth]{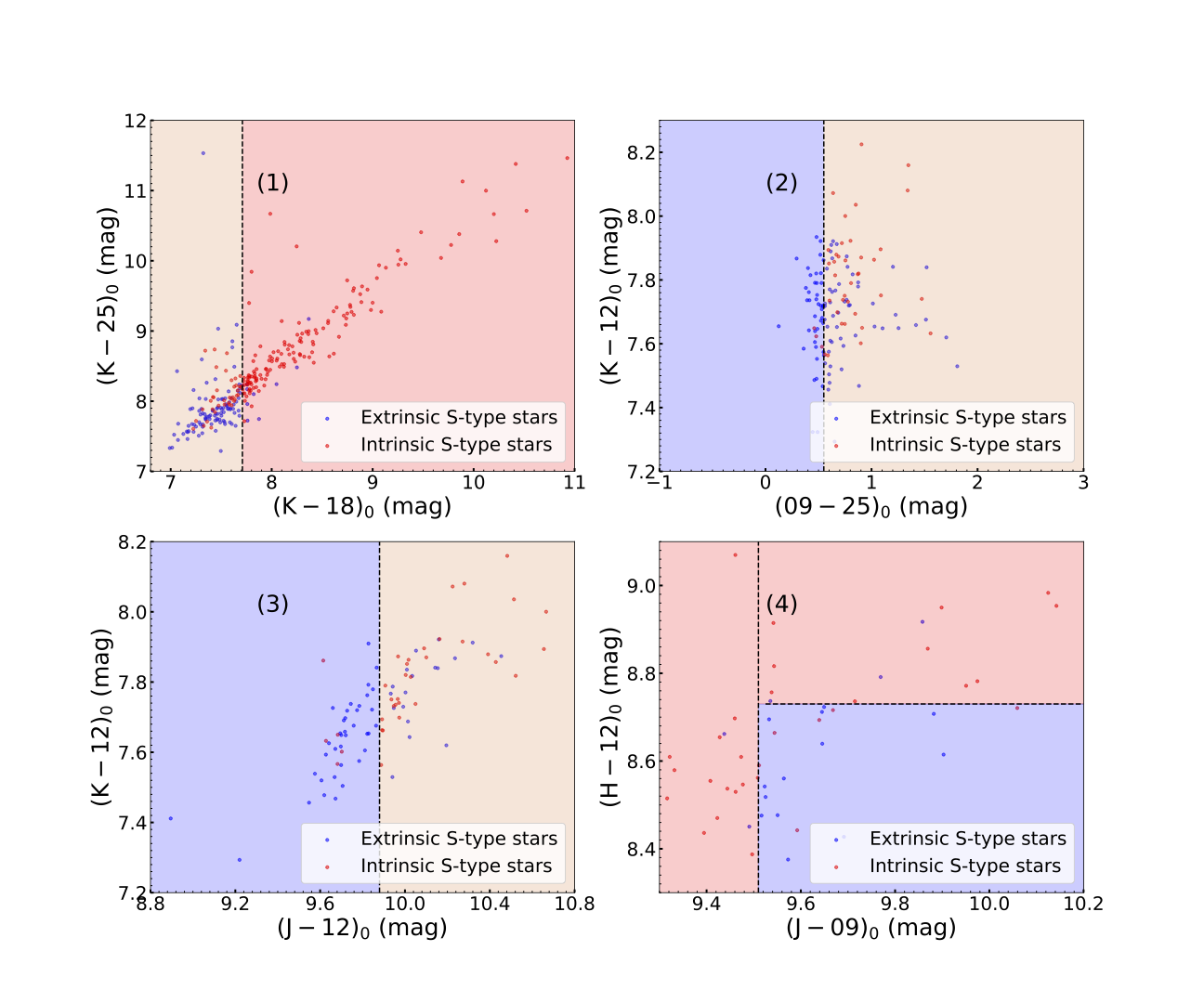}
	\caption{The infrared color criteria from the decision tree with the highest accuracy of classification between intrinsic and extrinsic stars. The red and blue dots denote intrinsic and extrinsic S-type stars collected from literature, respectively. The red and blue areas represent the regions of intrinsic and extrinsic S-type stars, respectively, and the yellow areas indicate stars in them need further classification with the following diagrams. The dashed vertical and horizontal lines represent the color criteria to classify intrinsic and extrinsic S-type stars in different color-color diagrams.
    \label{Color_criteria}} 
\end{figure*}

Using Fig.\ref{Decision_Tree}, we obtained more objective IR color-color criteria than those empirical ones in literature \citep{2019AJ....158...22C} to classify intrinsic and extrinsic S-type stars, and they were shown in Fig.\ref{Color_criteria}. The red and blue dots denote the intrinsic and extrinsic S-type stars mentioned in subsection \ref{sec:sample}, respectively, and the stars missing IR colors were removed because it is impossible to distinguish them using the color-color diagram method in this case. The red and blue areas represent the regions of intrinsic and extrinsic S-type stars, respectively, and the yellow areas indicate stars in them need further classification with the following diagrams. The dashed vertical and horizontal lines represent the color criteria to classify intrinsic and extrinsic S-type stars in different color-color diagrams, and they are listed as follows: 
\begin{gather}
   \mathrm{(K - [18])_0 > 7.71, \ for\ intrinsic\  stars} \\
   \mathrm{([09] - [25])_0 < 0.55, \ for\ extrinsic\ stars} \\
   \mathrm{(J - [12])_0 < 9.88,\ for\ extrinsic\ stars} \\
   \mathrm{(J - [09])_0 < 9.51, \ for\ intrinsic\ stars} \\
   \mathrm{(J - [09])_0 > 9.51\ and\ (H - [12])_0 > 8.73,}\\ \mathrm{for\ intrinsic\ stars} \nonumber\\
   \mathrm{(J - [09])_0 > 9.51\ and\ (H - [12])_0 < 8.73,}\\ \mathrm{for\ extrinsic\ stars} \nonumber
\end{gather}
In the 205 intrinsic and 151 extrinsic S-type stars in subsection \ref{sec:sample}, there are 183 intrinsic and 105 extrinsic S-type stars without missing IR colors, and 169 intrinsic and 91 extrinsic stars in them can be correctly classified by the above color-color criteria. Therefore, the accuracies of distinguishing intrinsic and extrinsic S-type stars by the above color criteria are 92.3\% and 86.7\%, respectively; it is better than the criteria in \cite{2019AJ....158...22C}, and their accuracy rate for selecting intrinsic S-type stars is about 83\%.

\subsection{The Two Methods Were Applied to Classify Intrinsic and Extrinsic Stars of the 2939 S-type Stars}\label{sec:classify}

The XGBoost model trained in subsection \ref{XGBoost_infraredcolor_results} and the color criteria in subsection \ref{sec:color_color_method} were used here to classify intrinsic and extrinsic S-type stars for the 2939 S-type stars of LAMOST DR10 found in subsection \ref{sec:lamost_dr10_sstar}. These stars were cross-matched with 2MASS, AKARI, IRAS, and Gaia DR3 within 5$''$, there are 876 common stars with parallaxes larger than 0, and 418 of them have all magnitudes or fluxes of the three IR surveys without missing values. The trained XGBoost model was used for all the 876 common stars since it can deal with the case of missing values, 381 intrinsic and 495 extrinsic S-type stars were finally classified, and they are marked as ``I" and ``E" in the ``$\mathrm{Class_X}$" column of Table \ref{tab:catalog}, respectively. The color criteria were applied to the 418 common stars without missing values, and 336 intrinsic and 82 extrinsic S-type stars were classified by this method, which was marked as ``I'' and ``E'' in the ``$\mathrm{Class_C}$" column of Table \ref{tab:catalog}, respectively. Among the 418 stars that can be classified by both the XGBoost model and the IR colors, 395 stars (318 intrinsic and 77 extrinsic S-type stars) of them have consistent types given by the two methods, but the other 23 stars have inconsistent types. The distinct classification results for the 23 stars are possibly caused by the discrepancy of the classification methods; the XGBoost model predicted types by taking the weighted average of 100 decision trees, while the color-color diagrams only used the most accurate one decision tree.

\begin{figure}[ht!] 
	\centering
        \includegraphics[width=0.5\textwidth]{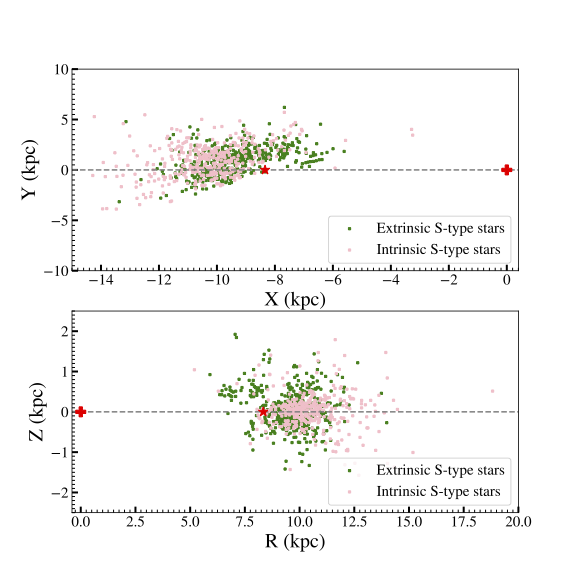}
	\caption{The spatial distribution of the 381 intrinsic and 495 extrinsic S-type stars, which were classified by the trained XGBoost model. The top panel shows the distribution on the galactic plane, and the bottom panel shows the distribution above or below the galactic disk. The green and pink dots represent extrinsic and intrinsic S-type stars, respectively, and the red star and filled plus denote the sun and Galactic center, respectively. \label{Gal_Distribution}}
\end{figure}

Figure \ref{Gal_Distribution} shows the spatial distribution of the 381 intrinsic and 495 extrinsic S-type stars classified by the trained XGBoost model, the distributions of intrinsic and extrinsic S-type stars are basically consistent, and most stars are located in the antigalactic center direction, especially the intrinsic S-type stars.

\section{Distinguish Intrinsic and Extrinsic S-type Stars Using Medium-Resolution Spectra of LAMOST}\label{sec:discuss}

In section \ref{sec:distinguish}, 381 intrinsic and 495 extrinsic S-type stars were classified by the XGBoost models from the 2939 S-type stars of LAMOST DR10, which provides enough training and testing samples, and makes the classification of intrinsic and extrinsic stars from spectra possible. In this section, we try to classify them using the MRS of LAMOST and the XGBoost model, and study which regions on spectra are most important to distinguish the two types of S -type stars.

\subsection{Use XGBoost Model to Classify Intrinsic and Extrinsic Stars from the 2939 S-type Stars}
\label{sec: spectra_xgboost}

In order to train and test the XGBoost classifier, the positive and negative samples, the training and testing data, and the input features needed to be determined. The LAMOST MRS of the 381 intrinsic and 495 extrinsic S-type stars classified by the trained XGBoost model in Section \ref{XGBoost_infraredcolor_results} was used to select positive and negative samples, which satisfy the criteria (1) the S/N of each spectrum is larger than 5; (2) low-quality spectra and spectra with extremely weak ZrO features were removed by manually checking; (3) if there is no blue- or red-band spectrum in exposure, this single exposure spectrum was discarded. Finally, 5675 and 5413 single-exposure and coadded spectra for the intrinsic and extrinsic S-type stars were selected as positive and negative samples, respectively, and these spectra were randomly divided into the training and testing samples in a ratio of 8:2. 

The blue-, red-, and blue+red- (combined) band spectra were used as input features to train the XGBoost model, respectively, and to test which case has the highest accuracy. The preprocessing process of all spectra is similar to that in section \ref{sec:lamost_spectra}, except that the blue-band spectra were included here, and each spectrum was rebinned with 1 $\rm \AA$ step.

When using the blue-band, red-band, and combined spectra as input features, the accuracy of the XGBoost model is 91.7\%, 93.6\%, and 94.8\%, respectively. Therefore, the combined spectra were finally used as input features to train XGBoost model, which has the highest accuracy, and 855 intrinsic and 2056 extrinsic stars in the 2939 S-type stars were classified by this method. However, there are still 28 S-type stars that have not been successfully classified because they do not have both blue- and red-band spectra in each exposure. The classification results using LAMOST spectra and XGBoost model were listed in the ``$\mathrm{Class_S}$" column of Table \ref{tab:catalog}, and the intrinsic and extrinsic S-type stars were marked as ``I'' and ``E,'' respectively.

In addition, we also checked whether the extrinsic S-type stars found in this work have obvious changes in RVs with the method as in Paper I. In order to avoid mixing nonbinary stars due to RV anomalies, the maximum value of RV differences ($\Delta$RV) between two exposures for each extrinsic star was removed, only the submaximal value of $\Delta$RV was used to select stars with obvious RV changes, and 526 of the 2939 stars with significant changes in RV were finally selected. Moreover, for the 495, 82, and 2056 extrinsic S-type stars classified by the three different methods in Sections \ref{sec:classify} and \ref{sec: spectra_xgboost}, there are 85, 13, and 383 stars with obvious changes in RV, respectively. The results are listed in the ``$\mathrm{E_{RV}}$" column of Table \ref{tab:catalog}, and the values of `` T " and `` F " indicate whether there is a significant change in RV, respectively.

\begin{figure*}[ht!] 
	\centering
        \includegraphics[width=1.0\textwidth]{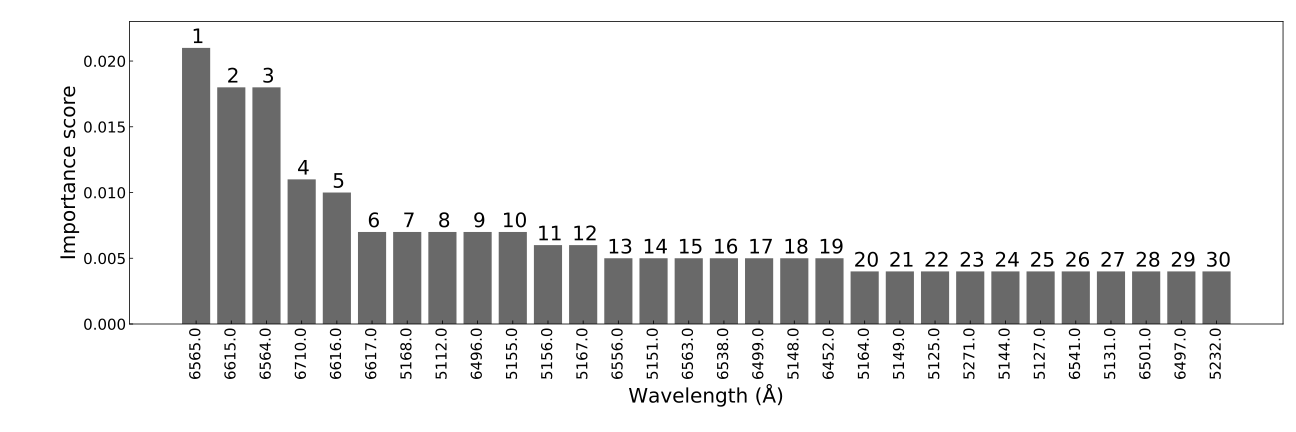}
	\caption{The top 30 most important features chosen by the XGBoost method, and the horizontal axis shows their wavelengths. \label{Spectrum_Importance}}
\end{figure*}

\subsection{The Significant Difference of Intrinsic and Extrinsic Stars in Spectra }

Intrinsic and extrinsic stars of the 2939 LAMOST S-type stars were successfully classified using MRS in the last subsection, and the main difference between them in the spectra was investigated here also using the XGBoost algorithm. The top 30 most important features of the training sample used by the trained XGBoost model in the last subsection were ranked in Fig.\ref{Spectrum_Importance}, and they can be used as a reference to find the main difference between the intrinsic and extrinsic S-type spectra. 

The nearby 10-20 $\mathrm{\AA}$ spectral region of each feature in Fig. \ref{Spectrum_Importance} was manually checked, and 13 spectral regions containing the 30 features were picked out. In these regions, the spectral lines containing the features in Fig.\ref{Spectrum_Importance} were used to study the difference between intrinsic and extrinsic stars on spectra. Nevertheless, if the nearby spectral lines have more obvious differences than the spectral lines containing the feature in Fig.\ref{Spectrum_Importance}, these nearby spectral lines would be used. After the 13 spectral lines in the 13 spectral regions were selected, their wavelength ranges and the pseudo-continuous spectra were defined, and their equivalent widths (EWs) for intrinsic and extrinsic stars in the training sample of the last subsection were calculated based on the following: 
\begin{gather}
    \mathrm{EW = \int_{\lambda_1}^{\lambda_2}(1-\frac{F_{I\lambda}}{F_{C\lambda}})d\lambda},
\end{gather}
where $\mathrm{F_{C\lambda}}$ and $\mathrm{F_{I\lambda}}$ are fluxes of the defined pseudo-continuous spectra and spectral lines in the 13 regions. For intrinsic and extrinsic stars in the training sample of the last subsection, we calculated the EWs of the 13 spectral lines for them, respectively, and found that there is a significant difference in the EWs of four spectral lines for them. 

\begin{figure*}[ht!] 
	\centering
        \includegraphics[width=1.0\textwidth]{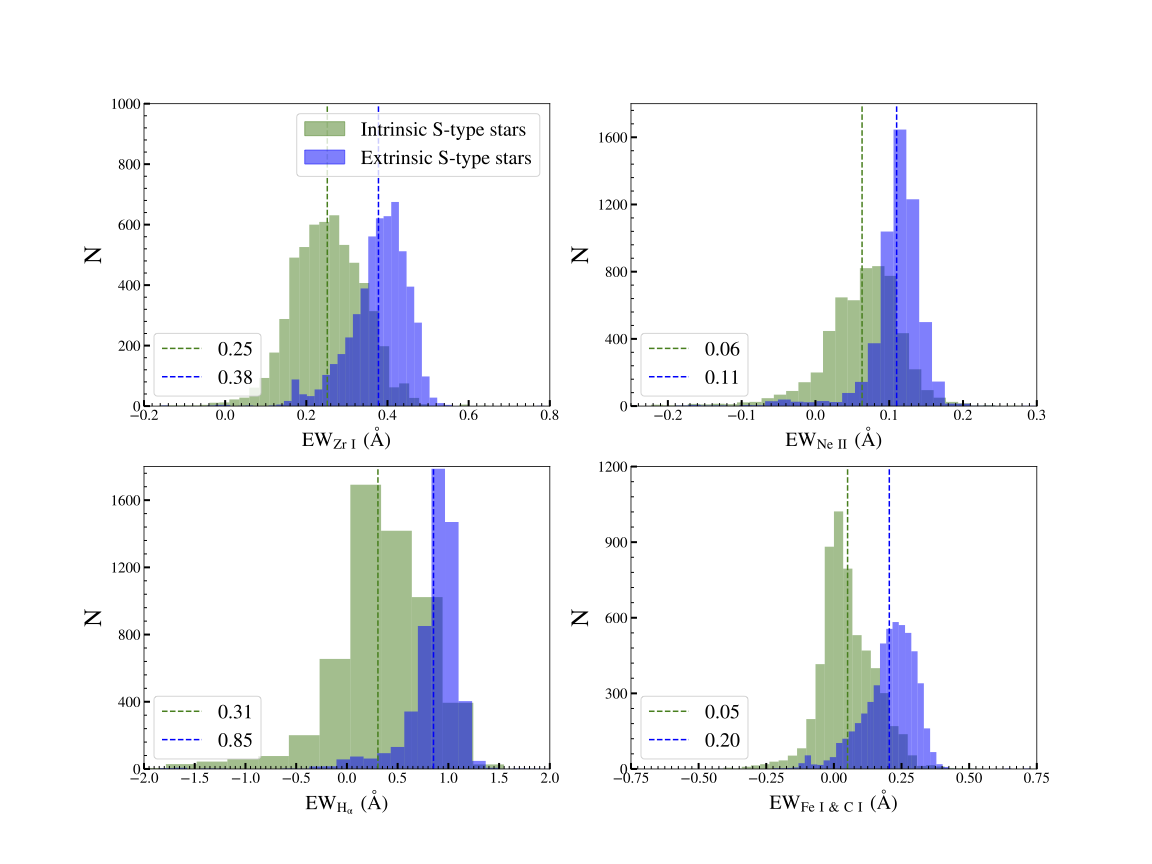}
	\caption{The equivalent widths of four important regions for intrinsic and extrinsic S-type stars. The dotted green and blue lines indicate the mean equivalent width for intrinsic and extrinsic S-type stars, respectively. \label{Spectrum_Classify}}
\end{figure*}

The defined wavelength ranges of the four spectral lines and the continua are listed in Table \ref{tab:line_band}, and its last column gives the serial numbers of spectral features of Fig.\ref{Spectrum_Importance} in the four spectral line regions. It should be noted that, for the fourth region (6609-6613 $\mathrm{\AA}$) in Table \ref{tab:line_band}, it does not contain any feature in Fig.\ref{Spectrum_Importance}, but next to the second, third, and sixth features. This is because the equivalent width of 6609-6613 $\mathrm{\AA}$ is obviously larger than the equivalent width of 6615-6617 $\mathrm{\AA}$ (containing the second, third, and sixth features), and the nearby region, 6609-6613 $\mathrm{\AA}$, of the second, third, and sixth features is finally selected to calculate EW as mentioned above. Their EW distributions of the four spectral lines were shown in the right panels of Fig.\ref{Spectrum_Classify}, and it can be seen that the median EWs of the four spectral lines for the extrinsic S-type stars is generally higher than that for the intrinsic S-type stars. \cite{2020AA...635L...6S} used MARCS model to predict the surface element abundances with atomic numbers from 36 to 82 (including Zr) for the intrinsic and extrinsic S-type stars, assuming that the intrinsic S-type stars have involved 5 pulses, and the companion stars of the extrinsic S-type stars have occurred in 61 pulses. They found that the abundances of the extrinsic S-type stars are generally higher than that of the intrinsic S-type stars, which is consistent with the previous results of Zr I EWs for intrinsic and extrinsic S-type stars, and this may be because the intrinsic S-type stars have not enough time to dredge-up elements to the surfaces of the stars. 

\begin{deluxetable*}{ccccc}
\setlength\tabcolsep{20pt}
\tablewidth{2000pt}
\caption{The Four Spectral Regions with Distinct Differences in the LAMOST MRS of the Intrinsic and Extrinsic S-type Stars and the Ranges of the Red-Blue Continuum Are Used to Calculate Their Equivalent Widths.}
\tablehead{
\colhead{$\mathrm{Index^a}$} & \colhead{Band ($\mathrm{\AA}$)} & \colhead{Blue Continuum ($\mathrm{\AA}$)} & \colhead{Red Continuum ($\mathrm{\AA}$)} & Serial Number in Figure \ref{Spectrum_Importance}
}
\startdata
Zr I & 6450.5-6452.5 & 6444.0-6447.0 & 6455.0-6457.0 & 19 \\
Ne II & 6538.5-6541.0 & 6536.0-6538.0 & 6541.5-6543.0 & 16 and 26\\
$\mathrm{H_{\alpha}}$ & 6563.0-6567.0 & 6560.0-6562.0 & 6568.0-6570.0 & 1, 3, and 15 \\
$\mathrm{Fe~I~\&~C~I}$ & 6609.0-6613.0 & 6602.0-6605.0 & 6618.0-6619.5 & near to 2, 5, and 6\\
\enddata
\tablecomments{$^a$:Denote the line index from NIST.}
\end{deluxetable*} \label{tab:line_band}

To test the effectiveness of the four regions for classifying intrinsic and extrinsic S-type stars, the positive and negative samples and the trained XGBoost model in last subsection were used, except that the input features were changed to the four spectrum regions, and they were reinterpolated to 0.1 $\mathrm{\AA}$. After testing, the accuracy of the model can reach to 92.1\%, and it is about 3\% lower than the previous accuracy (94.8\%) obtained with all the blue- and red-band spectra, so the four regions are important to distinguish between intrinsic and extrinsic S-type stars.

\section{Summary} \label{sec:summary}

In this paper, we provided a catalog of 2939 S-type stars; 2306 of them are reported as S-type stars for the first time, which were selected from the MRS of LAMOST DR10 by using two machine-learning methods. Based on IR photometric data, we classified these 2939 stars into intrinsic and extrinsic stars using both XGBoost model and color-color diagrams, respectively. Using these intrinsic and extrinsic S-type stars classified by IR photometric data as training samples, the XGBoost model was trained to reclassify 2939 stars by using spectroscopic data, and the regions with the most obvious spectral differences between intrinsic and extrinsic S-type stars were also investigated. The main conclusions of this work are as follows:
\begin{enumerate}
    \item We used the SVM and XGBoost algorithms to totally select 2939 S-type stars from LAMOST DR10, and further verify their nature of S-type stars. Using the 606 S-type stars of paper I, we trained the SVM and XGBoost models, manually examined nearly 130,000 spectra of S-type stars classified by the two methods, and found nearly 24,000 spectra (2939 stars) with obvious ZrO features. Excluding the stars reported in literature and papaer I, there are 2306 stars reported for the first time in this work. On the color-magnitude plot of $M_K$ vs. $W_{\mathrm{RP, BP-RP}} - W_{K, J-K}$, 764 S-type stars with high-quality Gaia data are indeed located in the O-rich AGB area. In addition, two methods were used to evaluate C/Os of 2939 stars, and they separately used the color-color diagrams constructed by the MARCS models of S-type stars and element abundances ([C/Fe] and [O/Fe]) of APOGEE DR17. The C/Os from the first method are in the range of [0.5, 1] as expected, and most of them are larger than 0.9. The C/Os of 18 stars from the second method are less than 0.5, but they were still contained in the S-type star catalog of this work because of the obvious ZrO features in their spectra. It should be noted that the C/O values from the two methods for individual stars differ significantly; this is because the uncertainties from a rough method used to estimate C/Os with color-color diagrams, the MARCS model themselves, and the algorithms used by APOGEE DR17; thus, the individual C/Os are not recommended to be used in this case.
    
    \item Two methods and IR colors were used to classify the 2939 stars into intrinsic and extrinsic S-type stars. Based on the known intrinsic and extrinsic S-type stars from literature and the IR data of 2MASS, AllWISE, AKARI, and IRAS, we analyzed the XGBoost model classification results using 15 input features based on IR data of the four sky surveys, and found that the XGBoost model trained by the input feature of 2AI, which used 2MASS, AKARI, and IRAS data, has the highest accuracy of 95.52\%. In this model, the two most important features are $K$ - [18] and [09] - [25], which may be caused by silicate signatures at 9.7 and 18 microns. Furthermore, the most accurate decision tree in the above trained XGBoost model was used to find a series of color criteria to classify intrinsic and extrinsic S-type stars, and the accuracy of these color criteria is about 90\%. There are 876 common stars (with missing values) in the 2939 S-type stars with 2MASS, AKARI, and IRAS; the trained XGBoost model was used to classify them into 381 intrinsic and 495 extrinsic S-type stars. In addition, there are 412 stars among the 876 common stars without missing IR data; they were classified into 336 intrinsic and 82 extrinsic S-type stars by using the IR color criteria, and of which 318 intrinsic and 77 extrinsic S-type stars have the consistent classification results with the XGBoost model.
    \item The 2939 S-type stars were classified into 855 intrinsic and 2056 extrinsic stars using their blue- and red-band MRS of LAMOST, and the most important spectral features to classify them were also investigated. Using the blue- and red-band spectra of the 381 intrinsic and 495 extrinsic S-type stars classified previously, we trained and tested XGBoost model, the accuracy to classify intrinsic and extrinsic stars from spectra is 94.82\%, and the 2939 stars were classified into 855 intrinsic and 2056 extrinsic stars with this model. Through analyzing the most important 30 spectral features used by this XGBoost model, 13 spectral regions containing the 30 features were selected, and 13 spectral lines and their continua in these regions were defined. After calculating the EWs of these spectral lines for the 381 intrinsic and 495 extrinsic stars, respectively, we found that the median EWs of four spectral regions of Zr I, Ne II, $\mathrm{H_{\alpha}}$, and Fe I $\&$ C I for extrinsic stars are significantly larger than those of the intrinsic stars. The EWs of Zr I of extrinsic stars are higher than that of intrinsic stars, which are consistent with the previous results \citep{2020AA...635L...6S}, and this may be because the intrinsic stars experience fewer pulses than the extrinsic stars. Thus, the four spectral regions were used to retrain and retest the XGBoost model, the accuracy to classify intrinsic and extrinsic stars can reach 92.1\%, which is about 3\% lower than the XGBoost model using both the blue- and red-band spectra, and they are really extremely important spectral features to distinguish between intrinsic and extrinsic S-type stars. In addition, after applying a similar method in Paper I to the 495, 82, and 2056 extrinsic stars classified by the three different methods, we found that 85, 13, and 383 extrinsic stars among them have obvious changes in RV, respectively.
\end{enumerate}

\section{Acknowledgements}
We thank Yanxin Guo, Yunjin Zhang, Xianglei Chen, and Jianjun Chen for useful discussions. This work is supported by National Science Foundation of China (grant Nos. U1931209, 12273078), China Manned Space Project ( Nos. CMS-CSST-2021-A10, CMS-CSST-2021-B05). Guoshoujing Telescope (the Large Sky Area Multi-Object Fiber Spectroscopic Telescope, LAMOST) is a National Major Scientific Project built by the Chinese Academy of Sciences. Funding for the Project has been provided by the National Development and Reform Commission. LAMOST is operated and managed by the National Astronomical Observatories, Chinese Academy of Sciences.
This research makes use of data from the European Space Agency (ESA) mission Gaia, processed by the Gaia Data Processing and Analysis Consortium. 

This research also makes use of Astropy, a community-developed core Python package for Astronomy \citep{2013A&A...558A..33A}, the TOPCAT tool \citep{2005ASPC..347...29T} and the VizieR catalog access tool and the Simbad database, operated at Centre de Donnees astronomiques de Strabourg (CDS), France.

$Facilities$: Du Pont (APOGEE), Sloan (APOGEE).

\bibliography{S_Type}{}

\begin{thebibliography}{}
\expandafter\ifx\csname natexlab\endcsname\relax\def\natexlab#1{#1}\fi
\providecommand{\url}[1]{\href{#1}{#1}}
\providecommand{\dodoi}[1]{doi:~\href{http://doi.org/#1}{\nolinkurl{#1}}}
\providecommand{\doeprint}[1]{\href{http://ascl.net/#1}{\nolinkurl{http://ascl.net/#1}}}
\providecommand{\doarXiv}[1]{\href{https://arxiv.org/abs/#1}{\nolinkurl{https://arxiv.org/abs/#1}}}

\bibitem[{{Abia} {et~al.}(2020){Abia}, {de Laverny}, {Cristallo}, {Kordopatis},
  \& {Straniero}}]{2020A&A...633A.135A}
{Abia}, C., {de Laverny}, P., {Cristallo}, S., {Kordopatis}, G., \&
  {Straniero}, O. 2020, \aap, 633, A135, \dodoi{10.1051/0004-6361/201936831}

\bibitem[{{Abia} {et~al.}(2022){Abia}, {de Laverny}, {Romero-G{\'o}mez}, \&
  {Figueras}}]{2022A&A...664A..45A}
{Abia}, C., {de Laverny}, P., {Romero-G{\'o}mez}, M., \& {Figueras}, F. 2022,
  \aap, 664, A45, \dodoi{10.1051/0004-6361/202243595}

\bibitem[{{Abrahamyan} {et~al.}(2015){Abrahamyan}, {Mickaelian}, \&
  {Knyazyan}}]{2015A&C....10...99A}
{Abrahamyan}, H.~V., {Mickaelian}, A.~M., \& {Knyazyan}, A.~V. 2015, Astronomy
  and Computing, 10, 99, \dodoi{10.1016/j.ascom.2014.12.002}

\bibitem[{{Alves}(2004)}]{2004NewAR..48..659A}
{Alves}, D.~R. 2004, \nar, 48, 659, \dodoi{10.1016/j.newar.2004.03.001}

\bibitem[{{Astropy Collaboration} {et~al.}(2013){Astropy Collaboration},
  {Robitaille}, {Tollerud}, {Greenfield}, {Droettboom}, {Bray}, {Aldcroft},
  {Davis}, {Ginsburg}, {Price-Whelan}, {Kerzendorf}, {Conley}, {Crighton},
  {Barbary}, {Muna}, {Ferguson}, {Grollier}, {Parikh}, {Nair}, {Unther},
  {Deil}, {Woillez}, {Conseil}, {Kramer}, {Turner}, {Singer}, {Fox}, {Weaver},
  {Zabalza}, {Edwards}, {Azalee Bostroem}, {Burke}, {Casey}, {Crawford},
  {Dencheva}, {Ely}, {Jenness}, {Labrie}, {Lim}, {Pierfederici}, {Pontzen},
  {Ptak}, {Refsdal}, {Servillat}, \& {Streicher}}]{2013A&A...558A..33A}
{Astropy Collaboration}, {Robitaille}, T.~P., {Tollerud}, E.~J., {et~al.} 2013,
  \aap, 558, A33, \dodoi{10.1051/0004-6361/201322068}

\bibitem[{{Beichman} {et~al.}(1988){Beichman}, {Neugebauer}, {Habing}, {Clegg},
  \& {Chester}}]{1988iras....1.....B}
{Beichman}, C.~A., {Neugebauer}, G., {Habing}, H.~J., {Clegg}, P.~E., \&
  {Chester}, T.~J. 1988, in Infrared astronomical satellite (IRAS) catalogs and
  atlases. Volume 1: Explanatory supplement, Vol.~1

\bibitem[{{Brewer} \& {Fischer}(2016)}]{2016ApJ...831...20B}
{Brewer}, J.~M., \& {Fischer}, D.~A. 2016, \apj, 831, 20,
  \dodoi{10.3847/0004-637X/831/1/20}

\bibitem[{{Brown} {et~al.}(1990){Brown}, {Smith}, {Lambert}, {Dutchover},
  {Hinkle}, \& {Johnson}}]{1990AJ.....99.1930B}
{Brown}, J.~A., {Smith}, V.~V., {Lambert}, D.~L., {et~al.} 1990, \aj, 99, 1930,
  \dodoi{10.1086/115475}

\bibitem[{{Busso} {et~al.}(1992){Busso}, {Gallino}, {Lambert}, {Raiteri}, \&
  {Smith}}]{1992ApJ...399..218B}
{Busso}, M., {Gallino}, R., {Lambert}, D.~L., {Raiteri}, C.~M., \& {Smith},
  V.~V. 1992, \apj, 399, 218, \dodoi{10.1086/171918}

\bibitem[{Chambers {et~al.}(2019)Chambers, Magnier, Metcalfe, Flewelling,
  Huber, Waters, Denneau, Draper, Farrow, Finkbeiner, Holmberg, Koppenhoefer,
  Price, Rest, Saglia, Schlafly, Smartt, Sweeney, Wainscoat, Burgett, Chastel,
  Grav, Heasley, Hodapp, Jedicke, Kaiser, Kudritzki, Luppino, Lupton, Monet,
  Morgan, Onaka, Shiao, Stubbs, Tonry, White, Bañados, Bell, Bender, Bernard,
  Boegner, Boffi, Botticella, Calamida, Casertano, Chen, Chen, Cole, Deacon,
  Frenk, Fitzsimmons, Gezari, Gibbs, Goessl, Goggia, Gourgue, Goldman, Grant,
  Grebel, Hambly, Hasinger, Heavens, Heckman, Henderson, Henning, Holman, Hopp,
  Ip, Isani, Jackson, Keyes, Koekemoer, Kotak, Le, Liska, Long, Lucey, Liu,
  Martin, Masci, McLean, Mindel, Misra, Morganson, Murphy, Obaika, Narayan,
  Nieto-Santisteban, Norberg, Peacock, Pier, Postman, Primak, Rae, Rai, Riess,
  Riffeser, Rix, Röser, Russel, Rutz, Schilbach, Schultz, Scolnic, Strolger,
  Szalay, Seitz, Small, Smith, Soderblom, Taylor, Thomson, Taylor, Thakar,
  Thiel, Thilker, Unger, Urata, Valenti, Wagner, Walder, Walter, Watters,
  Werner, Wood-Vasey, \& Wyse}]{chambers2019panstarrs1}
Chambers, K.~C., Magnier, E.~A., Metcalfe, N., {et~al.} 2019.
\newblock \doarXiv{1612.05560}

\bibitem[{{Chen} {et~al.}(2022){Chen}, {Luo}, {Li}, {Chen}, {Wang}, {Li}, {Du},
  \& {Ma}}]{2022ApJ...931..133C}
{Chen}, J., {Luo}, A.~L., {Li}, Y.-B., {et~al.} 2022, \apj, 931, 133,
  \dodoi{10.3847/1538-4357/ac66de}

\bibitem[{{Chen} {et~al.}(2019){Chen}, {Liu}, \& {Shan}}]{2019AJ....158...22C}
{Chen}, P.~S., {Liu}, J.~Y., \& {Shan}, H.~G. 2019, \aj, 158, 22,
  \dodoi{10.3847/1538-3881/ab2334}

\bibitem[{{Chen} {et~al.}(1998){Chen}, {Wang}, \&
  {Xiong}}]{1998A&A...333..613C}
{Chen}, P.~S., {Wang}, X.~H., \& {Xiong}, G.~Z. 1998, \aap, 333, 613

\bibitem[{{Chen} \& {Guestrin}(2016)}]{2016arXiv160302754C}
{Chen}, T., \& {Guestrin}, C. 2016, arXiv e-prints, arXiv:1603.02754.
\newblock \doarXiv{1603.02754}

\bibitem[{{Chiar} \& {Tielens}(2006)}]{2006ApJ...637..774C}
{Chiar}, J.~E., \& {Tielens}, A.~G.~G.~M. 2006, \apj, 637, 774,
  \dodoi{10.1086/498406}

\bibitem[{{Cohen} {et~al.}(2003){Cohen}, {Wheaton}, \&
  {Megeath}}]{2003AJ....126.1090C}
{Cohen}, M., {Wheaton}, W.~A., \& {Megeath}, S.~T. 2003, \aj, 126, 1090,
  \dodoi{10.1086/376474}

\bibitem[{{Cortes} \& {Vapnik}(1995)}]{1995ML...20....273}
{Cortes}, C., \& {Vapnik}, V. 1995, Machine Learning, 20, 273,
  \dodoi{10.1007/BF00994018}

\bibitem[{{Cui} {et~al.}(2012){Cui}, {Zhao}, {Chu}, {Li}, {Li}, {Zhang}, {Su},
  {Yao}, {Wang}, {Xing}, {Li}, {Zhu}, {Wang}, {Gu}, {Luo}, {Xu}, {Zhang},
  {Liu}, {Zhang}, {Yang}, {Cao}, {Chen}, {Chen}, {Chen}, {Chen}, {Chu}, {Feng},
  {Gong}, {Hou}, {Hu}, {Hu}, {Hu}, {Jia}, {Jiang}, {Jiang}, {Jiang}, {Jin},
  {Li}, {Li}, {Li}, {Liu}, {Liu}, {Lu}, {Mao}, {Men}, {Qi}, {Qi}, {Shi},
  {Tang}, {Tao}, {Wang}, {Wang}, {Wang}, {Wang}, {Wang}, {Wang}, {Wang},
  {Wang}, {Wang}, {Wang}, {Wang}, {Wang}, {Xu}, {Xu}, {Yang}, {Yu}, {Yuan},
  {Yuan}, {Zhai}, {Zhang}, {Zhang}, {Zhang}, {Zhao}, {Zhou}, {Zhou}, {Zhu}, \&
  {Zou}}]{2012RAA....12.1197C}
{Cui}, X.-Q., {Zhao}, Y.-H., {Chu}, Y.-Q., {et~al.} 2012, Research in Astronomy
  and Astrophysics, 12, 1197, \dodoi{10.1088/1674-4527/12/9/003}

\bibitem[{{Dolidze}(1975)}]{1975AbaOB..47....3D}
{Dolidze}, M.~V. 1975, Abastumanskaia Astrofizicheskaia Observatoriia
  Byulleten, 47, 3

\bibitem[{Fawcett(2006)}]{fawcett2006introduction}
Fawcett, T. 2006, Pattern Recognit. Lett, 27, 861

\bibitem[{{Fouesneau} {et~al.}(2022){Fouesneau}, {Andrae}, {Dharmawardena},
  {Rybizki}, {Bailer-Jones}, \& {Demleitner}}]{2022A&A...662A.125F}
{Fouesneau}, M., {Andrae}, R., {Dharmawardena}, T., {et~al.} 2022, \aap, 662,
  A125, \dodoi{10.1051/0004-6361/202141828}

\bibitem[{{Green} {et~al.}(2019){Green}, {Schlafly}, {Zucker}, {Speagle}, \&
  {Finkbeiner}}]{2019ApJ...887...93G}
{Green}, G.~M., {Schlafly}, E., {Zucker}, C., {Speagle}, J.~S., \&
  {Finkbeiner}, D. 2019, \apj, 887, 93, \dodoi{10.3847/1538-4357/ab5362}

\bibitem[{{Henize}(1960)}]{1960AJ.....65..491H}
{Henize}, K.~G. 1960, \aj, 65, 491, \dodoi{10.1086/108296}

\bibitem[{{Ishihara} {et~al.}(2010){Ishihara}, {Onaka}, {Kataza}, {Salama},
  {Alfageme}, {Cassatella}, {Cox}, {Garc{\'\i}a-Lario}, {Stephenson}, {Cohen},
  {Fujishiro}, {Fujiwara}, {Hasegawa}, {Ita}, {Kim}, {Matsuhara}, {Murakami},
  {M{\"u}ller}, {Nakagawa}, {Ohyama}, {Oyabu}, {Pyo}, {Sakon}, {Shibai},
  {Takita}, {Tanab{\'e}}, {Uemizu}, {Ueno}, {Usui}, {Wada}, {Watarai},
  {Yamamura}, \& {Yamauchi}}]{2010A&A...514A...1I}
{Ishihara}, D., {Onaka}, T., {Kataza}, H., {et~al.} 2010, \aap, 514, A1,
  \dodoi{10.1051/0004-6361/200913811}

\bibitem[{{Jarrett} {et~al.}(2011){Jarrett}, {Cohen}, {Masci}, {Wright},
  {Stern}, {Benford}, {Blain}, {Carey}, {Cutri}, {Eisenhardt}, {Lonsdale},
  {Mainzer}, {Marsh}, {Padgett}, {Petty}, {Ressler}, {Skrutskie}, {Stanford},
  {Surace}, {Tsai}, {Wheelock}, \& {Yan}}]{2011ApJ...735..112J}
{Jarrett}, T.~H., {Cohen}, M., {Masci}, F., {et~al.} 2011, \apj, 735, 112,
  \dodoi{10.1088/0004-637X/735/2/112}

\bibitem[{{Jin} {et~al.}(2019){Jin}, {Zhang}, {Zhang}, {Zhao}, {Wu}, \&
  {Fan}}]{2019MNRAS.485.4539J}
{Jin}, X., {Zhang}, Y., {Zhang}, J., {et~al.} 2019, \mnras, 485, 4539,
  \dodoi{10.1093/mnras/stz680}

\bibitem[{{Johnson}(1992)}]{1992IAUS..151..157J}
{Johnson}, H.~R. 1992, in Evolutionary Processes in Interacting Binary Stars,
  ed. Y.~{Kondo}, R.~{Sistero}, \& R.~S. {Polidan}, Vol. 151, 157

\bibitem[{{Jorissen} {et~al.}(2019){Jorissen}, {Boffin}, {Karinkuzhi}, {Van
  Eck}, {Escorza}, {Shetye}, \& {Van Winckel}}]{2019AA...626A.127J}
{Jorissen}, A., {Boffin}, H.~M.~J., {Karinkuzhi}, D., {et~al.} 2019, \aap, 626,
  A127, \dodoi{10.1051/0004-6361/201834630}

\bibitem[{{Jorissen} {et~al.}(1993){Jorissen}, {Frayer}, {Johnson}, {Mayor}, \&
  {Smith}}]{1993A&A...271..463J}
{Jorissen}, A., {Frayer}, D.~T., {Johnson}, H.~R., {Mayor}, M., \& {Smith},
  V.~V. 1993, \aap, 271, 463

\bibitem[{{Jorissen} \& {Mayor}(1988)}]{1988A&A...198..187J}
{Jorissen}, A., \& {Mayor}, M. 1988, \aap, 198, 187

\bibitem[{{Jorissen} \& {Mayor}(1992)}]{1992A&A...260..115J}
---. 1992, \aap, 260, 115

\bibitem[{{Keenan} \& {Boeshaar}(1980)}]{1980ApJS...43..379K}
{Keenan}, P.~C., \& {Boeshaar}, P.~C. 1980, \apjs, 43, 379,
  \dodoi{10.1086/190673}

\bibitem[{{Lebzelter} {et~al.}(2018){Lebzelter}, {Mowlavi}, {Marigo},
  {Pastorelli}, {Trabucchi}, {Wood}, \&
  {Lecoeur-Ta{\"\i}bi}}]{2018A&A...616L..13L}
{Lebzelter}, T., {Mowlavi}, N., {Marigo}, P., {et~al.} 2018, \aap, 616, L13,
  \dodoi{10.1051/0004-6361/201833615}

\bibitem[{{Li} {et~al.}(2018){Li}, {Luo}, {Du}, {Zuo}, {Wang}, {Zhao}, {Jiang},
  {Zhang}, {Liu}, {Qin}, {Wang}, {Du}, {Guo}, {Wang}, {Han}, {Xiang}, {Huang},
  {Chen}, {Chen}, {Kong}, {Hou}, {Song}, {Wang}, {Wu}, {Zhang}, {Zhang},
  {Wang}, {Cao}, {Hou}, \& {Zhao}}]{2018ApJS..234...31L}
{Li}, Y.-B., {Luo}, A.~L., {Du}, C.-D., {et~al.} 2018, \apjs, 234, 31,
  \dodoi{10.3847/1538-4365/aaa415}

\bibitem[{{Lindegren} {et~al.}(2018){Lindegren}, {Hern{\'a}ndez}, {Bombrun},
  {Klioner}, {Bastian}, {Ramos-Lerate}, {de Torres}, {Steidelm{\"u}ller},
  {Stephenson}, {Hobbs}, {Lammers}, {Biermann}, {Geyer}, {Hilger}, {Michalik},
  {Stampa}, {McMillan}, {Casta{\~n}eda}, {Clotet}, {Comoretto}, {Davidson},
  {Fabricius}, {Gracia}, {Hambly}, {Hutton}, {Mora}, {Portell}, {van Leeuwen},
  {Abbas}, {Abreu}, {Altmann}, {Andrei}, {Anglada}, {Balaguer-N{\'u}{\~n}ez},
  {Barache}, {Becciani}, {Bertone}, {Bianchi}, {Bouquillon}, {Bourda},
  {Br{\"u}semeister}, {Bucciarelli}, {Busonero}, {Buzzi}, {Cancelliere},
  {Carlucci}, {Charlot}, {Cheek}, {Crosta}, {Crowley}, {de Bruijne}, {de
  Felice}, {Drimmel}, {Esquej}, {Fienga}, {Fraile}, {Gai}, {Garralda},
  {Gonz{\'a}lez-Vidal}, {Guerra}, {Hauser}, {Hofmann}, {Holl}, {Jordan},
  {Lattanzi}, {Lenhardt}, {Liao}, {Licata}, {Lister}, {L{\"o}ffler},
  {Marchant}, {Martin-Fleitas}, {Messineo}, {Mignard}, {Morbidelli}, {Poggio},
  {Riva}, {Rowell}, {Salguero}, {Sarasso}, {Sciacca}, {Siddiqui}, {Smart},
  {Spagna}, {Steele}, {Taris}, {Torra}, {van Elteren}, {van Reeven}, \&
  {Vecchiato}}]{2018A&A...616A...2L}
{Lindegren}, L., {Hern{\'a}ndez}, J., {Bombrun}, A., {et~al.} 2018, \aap, 616,
  A2, \dodoi{10.1051/0004-6361/201832727}

\bibitem[{{Liu} {et~al.}(2020){Liu}, {Fu}, {Shi}, {Wu}, {Han}, {Chen}, {Dong},
  {Zhao}, {Chen}, {Zhang}, {Bai}, {Chen}, {Cui}, {Du}, {Hsia}, {Jiang}, {Hou},
  {Hou}, {Li}, {Li}, {Li}, {Liu}, {Liu}, {Luo}, {Ren}, {Tian}, {Tian}, {Wang},
  {Wu}, {Xie}, {Yan}, {Yang}, {Yu}, {Zhang}, {Zhang}, {Zhang}, {Zhang}, {Zhao},
  {Zhong}, {Zong}, \& {Zuo}}]{2020arXiv200507210L}
{Liu}, C., {Fu}, J., {Shi}, J., {et~al.} 2020, arXiv e-prints,
  arXiv:2005.07210.
\newblock \doarXiv{2005.07210}

\bibitem[{{Luo} {et~al.}(2012){Luo}, {Zhang}, {Zhao}, {Zhao}, {Cui}, {Li},
  {Chu}, {Shi}, {Wang}, {Zhang}, {Bai}, {Chen}, {Wang}, {Guo}, {Chen}, {Du},
  {Kong}, {Lei}, {Li}, {Song}, {Wu}, {Zhang}, {Zhou}, {Zuo}, {Du}, {He}, {Hou},
  {Dong}, {Li}, {Li}, {Li}, {Song}, {Tian}, {Wang}, {Wu}, {Yang}, {Yuan},
  {Cao}, {Chen}, {Chen}, {Chen}, {Chu}, {Feng}, {Gong}, {Gu}, {Hou}, {Huo},
  {Hu}, {Hu}, {Hu}, {Jia}, {Jiang}, {Jiang}, {Jiang}, {Jin}, {Li}, {Li}, {Li},
  {Li}, {Li}, {Liu}, {Liu}, {Liu}, {Lu}, {Lu}, {Luo}, {Mao}, {Men}, {Ni}, {Qi},
  {Qi}, {Shi}, {Su}, {Sun}, {Su}, {Tang}, {Tao}, {Tu}, {Wang}, {Wang}, {Wang},
  {Wang}, {Wang}, {Wang}, {Wang}, {Wang}, {Wang}, {Wang}, {Wang}, {Wang},
  {Wang}, {Wang}, {Wei}, {Xue}, {Xing}, {Xu}, {Xu}, {Xu}, {Yang}, {Yang},
  {Yao}, {Yu}, {Yuan}, {Zhai}, {Zhang}, {Zhang}, {Zhang}, {Zhang}, {Zhang},
  {Zhang}, {Zhao}, {Zhou}, {Zhu}, {Zhu}, \& {Zou}}]{2012RAA....12.1243L}
{Luo}, A.~L., {Zhang}, H.-T., {Zhao}, Y.-H., {et~al.} 2012, Research in
  Astronomy and Astrophysics, 12, 1243, \dodoi{10.1088/1674-4527/12/9/004}

\bibitem[{{MacConnell}(1979)}]{1979A&AS...38..335M}
{MacConnell}, D.~J. 1979, \aaps, 38, 335

\bibitem[{{Marchetti} {et~al.}(2022){Marchetti}, {Evans}, \&
  {Rossi}}]{2022MNRAS.515..767M}
{Marchetti}, T., {Evans}, F.~A., \& {Rossi}, E.~M. 2022, \mnras, 515, 767,
  \dodoi{10.1093/mnras/stac1777}

\bibitem[{{Marchetti} {et~al.}(2019){Marchetti}, {Rossi}, \&
  {Brown}}]{2019MNRAS.490..157M}
{Marchetti}, T., {Rossi}, E.~M., \& {Brown}, A.~G.~A. 2019, \mnras, 490, 157,
  \dodoi{10.1093/mnras/sty2592}

\bibitem[{{Merrill}(1952)}]{1952ApJ...116...21M}
{Merrill}, P.~W. 1952, \apj, 116, 21, \dodoi{10.1086/145589}

\bibitem[{{Murakami} {et~al.}(2007){Murakami}, {Baba}, {Barthel}, {Clements},
  {Cohen}, {Doi}, {Enya}, {Figueredo}, {Fujishiro}, {Fujiwara}, {Fujiwara},
  {Garcia-Lario}, {Goto}, {Hasegawa}, {Hibi}, {Hirao}, {Hiromoto}, {Hong},
  {Imai}, {Ishigaki}, {Ishiguro}, {Ishihara}, {Ita}, {Jeong}, {Jeong},
  {Kaneda}, {Kataza}, {Kawada}, {Kawai}, {Kawamura}, {Kessler}, {Kester},
  {Kii}, {Kim}, {Kim}, {Kobayashi}, {Koo}, {Kwon}, {Lee}, {Lorente}, {Makiuti},
  {Matsuhara}, {Matsumoto}, {Matsuo}, {Matsuura}, {M{\"U}ller}, {Murakami},
  {Nagata}, {Nakagawa}, {Naoi}, {Narita}, {Noda}, {Oh}, {Ohnishi}, {Ohyama},
  {Okada}, {Okuda}, {Oliver}, {Onaka}, {Ootsubo}, {Oyabu}, {Pak}, {Park},
  {Pearson}, {Rowan-Robinson}, {Saito}, {Sakon}, {Salama}, {Sato}, {Savage},
  {Serjeant}, {Shibai}, {Shirahata}, {Sohn}, {Suzuki}, {Takagi}, {Takahashi},
  {Tanab{\'E}}, {Takeuchi}, {Takita}, {Thomson}, {Uemizu}, {Ueno}, {Usui},
  {Verdugo}, {Wada}, {Wang}, {Watabe}, {Watarai}, {White}, {Yamamura},
  {Yamauchi}, \& {Yasuda}}]{2007PASJ...59S.369M}
{Murakami}, H., {Baba}, H., {Barthel}, P., {et~al.} 2007, \pasj, 59, S369,
  \dodoi{10.1093/pasj/59.sp2.S369}

\bibitem[{{Otto} {et~al.}(2011){Otto}, {Green}, \&
  {Gray}}]{2011ApJS..196....5O}
{Otto}, E., {Green}, P.~J., \& {Gray}, R.~O. 2011, \apjs, 196, 5,
  \dodoi{10.1088/0067-0049/196/1/5}

\bibitem[{{Shetye} {et~al.}(2019){Shetye}, {Goriely}, {Siess}, {Van Eck},
  {Jorissen}, \& {Van Winckel}}]{2019AA...625L...1S}
{Shetye}, S., {Goriely}, S., {Siess}, L., {et~al.} 2019, \aap, 625, L1,
  \dodoi{10.1051/0004-6361/201935296}

\bibitem[{{Shetye} {et~al.}(2020){Shetye}, {Van Eck}, {Goriely}, {Siess},
  {Jorissen}, {Escorza}, \& {Van Winckel}}]{2020AA...635L...6S}
{Shetye}, S., {Van Eck}, S., {Goriely}, S., {et~al.} 2020, \aap, 635, L6,
  \dodoi{10.1051/0004-6361/202037481}

\bibitem[{{Shetye} {et~al.}(2018){Shetye}, {Van Eck}, {Jorissen}, {Van
  Winckel}, {Siess}, {Goriely}, {Escorza}, {Karinkuzhi}, \&
  {Plez}}]{2018AA...620A.148S}
{Shetye}, S., {Van Eck}, S., {Jorissen}, A., {et~al.} 2018, \aap, 620, A148,
  \dodoi{10.1051/0004-6361/201833298}

\bibitem[{{Shetye} {et~al.}(2021){Shetye}, {Van Eck}, {Jorissen}, {Goriely},
  {Siess}, {Van Winckel}, {Plez}, {Godefroid}, \&
  {Wallerstein}}]{2021AA...650A.118S}
---. 2021, \aap, 650, A118, \dodoi{10.1051/0004-6361/202040207}

\bibitem[{{Skrutskie} {et~al.}(2006){Skrutskie}, {Cutri}, {Stiening},
  {Weinberg}, {Schneider}, {Carpenter}, {Beichman}, {Capps}, {Chester},
  {Elias}, {Huchra}, {Liebert}, {Lonsdale}, {Monet}, {Price}, {Seitzer},
  {Jarrett}, {Kirkpatrick}, {Gizis}, {Howard}, {Evans}, {Fowler}, {Fullmer},
  {Hurt}, {Light}, {Kopan}, {Marsh}, {McCallon}, {Tam}, {Van Dyk}, \&
  {Wheelock}}]{2006AJ....131.1163S}
{Skrutskie}, M.~F., {Cutri}, R.~M., {Stiening}, R., {et~al.} 2006, \aj, 131,
  1163, \dodoi{10.1086/498708}

\bibitem[{{Soszy{\'n}ski} {et~al.}(2005){Soszy{\'n}ski}, {Gieren}, \&
  {Pietrzy{\'n}ski}}]{2005PASP..117..823S}
{Soszy{\'n}ski}, I., {Gieren}, W., \& {Pietrzy{\'n}ski}, G. 2005, \pasp, 117,
  823, \dodoi{10.1086/431434}

\bibitem[{{Stephenson}(1984)}]{1984PW&SO...3....1S}
{Stephenson}, C.~B. 1984, Publications of the Warner \& Swasey Observatory

\bibitem[{{Su} \& {Cui}(2004)}]{2004ChJAA...4....1S}
{Su}, D.-Q., \& {Cui}, X.-Q. 2004, \cjaa, 4, 1, \dodoi{10.1088/1009-9271/4/1/1}

\bibitem[{{Taylor}(2005)}]{2005ASPC..347...29T}
{Taylor}, M.~B. 2005, in Astronomical Society of the Pacific Conference Series,
  Vol. 347, Astronomical Data Analysis Software and Systems XIV, ed.
  P.~{Shopbell}, M.~{Britton}, \& R.~{Ebert}, 29

\bibitem[{{Van Eck} \& {Jorissen}(1999)}]{1999A&A...345..127V}
{Van Eck}, S., \& {Jorissen}, A. 1999, \aap, 345, 127.
\newblock \doarXiv{astro-ph/9903241}

\bibitem[{{Van Eck} {et~al.}(2022){Van Eck}, {Shetye}, \&
  {Siess}}]{2022Univ....8..220V}
{Van Eck}, S., {Shetye}, S., \& {Siess}, L. 2022, Universe, 8, 220,
  \dodoi{10.3390/universe8040220}

\bibitem[{{Van Eck} {et~al.}(2017){Van Eck}, {Neyskens}, {Jorissen}, {Plez},
  {Edvardsson}, {Eriksson}, {Gustafsson}, {J{\o}rgensen}, \&
  {Nordlund}}]{2017A&A...601A..10V}
{Van Eck}, S., {Neyskens}, P., {Jorissen}, A., {et~al.} 2017, \aap, 601, A10,
  \dodoi{10.1051/0004-6361/201525886}

\bibitem[{{Wang} {et~al.}(1996){Wang}, {Su}, {Chu}, {Cui}, \&
  {Wang}}]{1996ApOpt..35.5155W}
{Wang}, S.-G., {Su}, D.-Q., {Chu}, Y.-Q., {Cui}, X., \& {Wang}, Y.-N. 1996,
  \ao, 35, 5155, \dodoi{10.1364/AO.35.005155}

\bibitem[{{Wang} \& {Chen}(2002)}]{2002A&A...387..129W}
{Wang}, X.~H., \& {Chen}, P.~S. 2002, \aap, 387, 129,
  \dodoi{10.1051/0004-6361:20020356}

\bibitem[{{Wright} {et~al.}(2010){Wright}, {Eisenhardt}, {Mainzer}, {Ressler},
  {Cutri}, {Jarrett}, {Kirkpatrick}, {Padgett}, {McMillan}, {Skrutskie},
  {Stanford}, {Cohen}, {Walker}, {Mather}, {Leisawitz}, {Gautier}, {McLean},
  {Benford}, {Lonsdale}, {Blain}, {Mendez}, {Irace}, {Duval}, {Liu}, {Royer},
  {Heinrichsen}, {Howard}, {Shannon}, {Kendall}, {Walsh}, {Larsen}, {Cardon},
  {Schick}, {Schwalm}, {Abid}, {Fabinsky}, {Naes}, \&
  {Tsai}}]{2010AJ....140.1868W}
{Wright}, E.~L., {Eisenhardt}, P. R.~M., {Mainzer}, A.~K., {et~al.} 2010, \aj,
  140, 1868, \dodoi{10.1088/0004-6256/140/6/1868}

\bibitem[{{Yan} {et~al.}(2022){Yan}, {Li}, {Wang}, {Zong}, {Yuan}, {Xiang},
  {Huang}, {Xie}, {Dong}, {Yuan}, {Bi}, {Chu}, {Cui}, {Deng}, {Fu}, {Han},
  {Hou}, {Li}, {Liu}, {Liu}, {Liu}, {Luo}, {Shi}, {Wu}, {Zhang}, {Zhao}, \&
  {Zhao}}]{2022Innov...300224Y}
{Yan}, H., {Li}, H., {Wang}, S., {et~al.} 2022, The Innovation, 3, 100224,
  \dodoi{10.1016/j.xinn.2022.100224}

\bibitem[{{Yang} {et~al.}(2006){Yang}, {Chen}, {Wang}, \&
  {He}}]{2006AJ....132.1468Y}
{Yang}, X., {Chen}, P., {Wang}, J., \& {He}, J. 2006, \aj, 132, 1468,
  \dodoi{10.1086/506965}

\bibitem[{{Yi} {et~al.}(2019){Yi}, {Chen}, {Pan}, {Yue}, {Lu}, {Li}, \&
  {Luo}}]{2019ApJ...887..241Y}
{Yi}, Z., {Chen}, Z., {Pan}, J., {et~al.} 2019, \apj, 887, 241,
  \dodoi{10.3847/1538-4357/ab54d0}

\bibitem[{{Zhao} \& {Newberg}(2006)}]{2006astro.ph.12034Z}
{Zhao}, C., \& {Newberg}, H.~J. 2006, arXiv e-prints, astro.
\newblock \doarXiv{astro-ph/0612034}

\bibitem[{{Zhao} {et~al.}(2012){Zhao}, {Zhao}, {Chu}, {Jing}, \&
  {Deng}}]{2012RAA....12..723Z}
{Zhao}, G., {Zhao}, Y.-H., {Chu}, Y.-Q., {Jing}, Y.-P., \& {Deng}, L.-C. 2012,
  Research in Astronomy and Astrophysics, 12, 723,
  \dodoi{10.1088/1674-4527/12/7/002}

\end{thebibliography}
\bibliographystyle{aasjournal}

\end{document}